\documentclass{jfm}
\usepackage{amsmath}
\usepackage[mathlines]{lineno}
\makeatletter
\let\DOTSI\relax 
\newcommand*{\Uint}{%
  \DOTSI
  \mathop{%
    \mathpalette\@LetterOnInt{\smile}%
  }%
  \mkern-\thinmuskip 
  \int
}
\newcommand*{\@LetterOnInt}[2]{%
  \sbox0{$#1\int\m@th$}%
  \sbox2{$%
    \ifx#1\displaystyle
      \textstyle
    \else
      \scriptscriptstyle
    \fi
    #2%
  \m@th$}%
  \dimen@=.4\dimexpr\ht0+\dp0\relax
  \ifdim\dimexpr\ht2+\dp2\relax>\dimen@
    \sbox2{\resizebox*{!}{\dimen@}{\unhcopy2}}%
  \fi
  \dimen@=\wd0 %
  \ifdim\wd2>\dimen@
    \dimen@=\wd2 %
  \fi
  \rlap{\hbox to \dimen@{\hfil
    $#1\vcenter{\copy2}\m@th$%
  \hfil}}%
  \ifdim\dimen@>\wd0 %
    \kern.5\dimexpr\dimen@-\wd0\relax
  \fi
}
\makeatother

\usepackage[dvipsnames]{xcolor}
\newcommand{\change}[1]{#1}
\newcommand{\ctwo}[1]{#1}

\newcommand{\e}{\mathrm{e}}
\newcommand{\sech}{\mathrm{sech}}
\newcommand{\upd}{\mathrm{d}}
\renewcommand{\upi}{\mathrm{i}}
\renewcommand{\Re}{\mathrm{Re}}

\newcommand{\R}{\mathbb{R}}

\newcommand{\w}{\mathrm{w}}

\shorttitle{Wilks et al.}
\shortauthor{B. Wilks, M. H. Meylan, Z. J. Wegert, V. J. Challis \& N. Thamwattana}

\title{Water wave scattering by a surface-mounted rectangular anisotropic elastic plate}

\author{Ben Wilks\aff{1,2}\corresp{\email{\ctwo{b.wilks@adelaide.edu.au}}}, Michael H. Meylan\aff{1}, Zachary J. Wegert\aff{3}, Vivien J. Challis\aff{3} \and Ngamta Thamwattana\aff{1}}

\affiliation{\aff{1}School of Computer and Information Sciences, University of Newcastle, Callaghan NSW 2308, Australia
\aff{2} \ctwo{School of Mathematical Sciences, Adelaide University}, Mawson Lakes SA 5095, Australia
\aff{3}School of Mathematical Sciences, Queensland University of Technology, Brisbane QLD 4001, Australia}

\begin{document}
\maketitle

\begin{abstract}
\change{This paper considers the problem of water wave scattering by a rectangular anisotropic elastic plate mounted on the ocean surface, with either \ctwo{free, clamped or simply-supported} edges. The problem is obtained as an expansion over the dry modes of the elastic plate, which are computed using a Rayleigh--Ritz method.} In turn, the component diffraction and radiation problems are solved by formulating a boundary integral equation and solving numerically using a constant panel method. The results are presented to highlight the resonant responses of the plate under different forcing scenarios. In particular, we illustrate how the excitation of certain modes can be forbidden due to symmetry.
\end{abstract}


\section{Introduction}

Hydroelasticity, which is concerned with the interaction of fluids with elastic bodies, is an important topic with numerous applications, including those in the cryosphere (ocean wave scattering by sea ice and ice shelves) and those related to the engineering of very large floating structures (VLFS). A comprehensive account of the historical research can be found in the review articles \cite{squire2007ocean, squire2020ocean}. We give here a brief account of some of the hydroelastic models. We note that this extensive research has focussed almost entirely on isotropic plates\ctwo{, with a notable exception being the recent work of \citet{THERY2026103642}, who used a space transformation approach to determine the parameters of an anisotropic plate mounted over shallow water for cloaking applications}.

The two-dimensional water wave problem (one horizontal and one vertical) was solved by an eigenfunction expansion method by \cite{fox_squire94} and extended by \cite{sahoo2001scattering} who analysed a semi-infinite elastic plate floating on the surface of water of finite-depth. The Wiener-Hopf technique is particularly suited to semifinite problems and highly analytic solutions have been derived in this case \citep{balmforth1999ocean, chung2002calculation, tkacheva2003plane, williams_meylan12, smith2020wiener}. The finite elastic plate was solved by \cite{meylan_squire94} using Green's function technique, which led to a Fredholm integral equation and was simultaneously solved using a modal expansion by \cite{newman1994wave}. The three-dimensional water wave problem involving circular elastic discs was solved by \cite{jgrfloecirc,peter_meylan_chung04}. The solution for plates of arbitrary shapes was given in \cite{AppliedOR01,meylan2002wave}. Similar methods were developed by \cite{hermans2004interaction,andrianov2006hydroelastic}.


Recently, extensive studies on hydroelasticity have been driven by its potential applications in wave energy conversion, with flexible wave energy converters (WECs) continuing to be a focal point of current research \citep{Babarit2017,Philen2018,Tay2020,Collins2021,Renzi2021, Michele2022,Zheng2022,Vipin_Koley_2022,Guo2023,singh2023performance,Cheng2024,MEYLAN2025120931}. A promising mechanism for flexible wave energy conversion is the piezoelectric effect, in which materials become electrically polarised in response to elastic deformation. First comprehensively modelled by \citet{renzi2016hydroelectromechanical}, a piezoelectric WEC (usually modelled as a plate) would become polarised in response to deformation by ocean waves and thus induce a current in an external circuit. An important aspect of piezoelectric materials that has been hitherto ignored in the piezoelectric WEC literature is their anisotropy---while piezoceramics are usually transversely isotropic \citep{Yang_2018}, the polymer PVDF, which is often proposed for piezoelectric WECs, are anisotropic \citep{Vinogradov01041999}. To date, authors have bypassed considerations of anisotropy by focussing on the one-dimensional WEC/two-dimensional fluid context, with submerged horizontal plates \citep{renzi2016hydroelectromechanical,Vipin_Koley_2022}, surface mounted plates \citep{MEYLAN2025120931} and vertical plates \citep{singh2023performance} being among the cases considered. The present paper seeks to investigate wave scattering by a two-dimensional anisotropic elastic plate in a three-dimensional fluid. In doing so, this paper will facilitate future extensions of earlier models of piezoelectric WECs \citep{renzi2016hydroelectromechanical,MEYLAN2025120931} to three dimensions.

The outline of this paper is as follows. In \textsection\ref{outline sec} we describe a model of a rectangular anisotropic elastic plate mounted on the water surface. Our method is an extension of that presented by \citet{MEYLAN2025120931} for the two-dimensional case, in which the solution is expanded as a sum of a diffraction potential and a superposition of radiation potentials. The diffraction potential, which assumes that the plate is fixed, is solved in \textsection\ref{rigid_plate_sec}. The modes of vibration of the plate and the corresponding radiation potentials are found in \textsection\ref{vibration_modes_sec} and \textsection\ref{radiation_potentials_sec}, respectively. The coupled fluid/plate problem is solved in \textsection\ref{coupled_sec}. An energy balance identity used to verify our computations, which is based on the optical theorem, is derived in \textsection\ref{energy_balance_sec}. \change{A large number of results are presented in \textsection\ref{results_sec}, which serve to illustrate the capabilities of the model and act as benchmark solutions for future studies. A} conclusion is given in \textsection\ref{conclusion_sec}.

\section{Problem outline}\label{outline sec}
\subsection{Preliminaries}
\begin{figure}
    \centering
    \includegraphics[width=\linewidth]{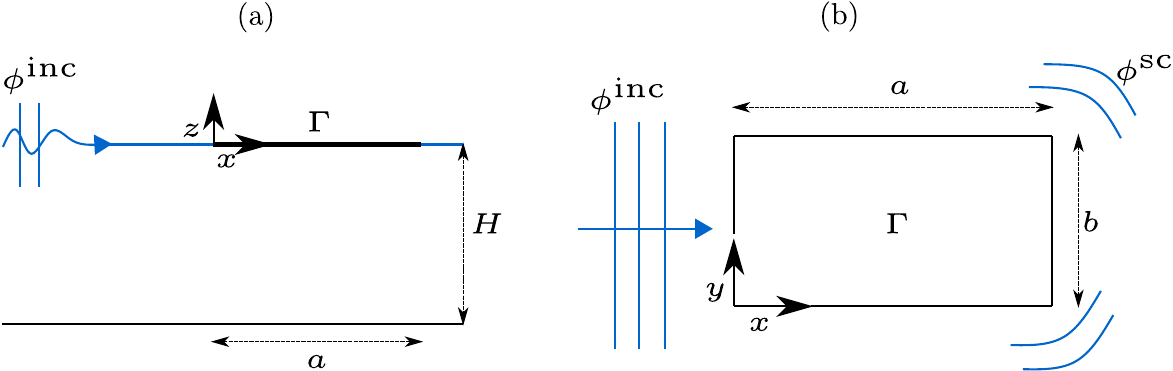}
    \caption{\change{(a) Side and (b) plan views of the scattering problem. The rectangular plate, labelled $\Gamma$, has side lengths $a$ and $b$ and the fluid is of depth $H$.} The incident wave $\phi^{\rm inc}$ excites the plate into motion, generating scattered waves $\phi^{\rm sc}$. }
    \label{fig:schematic}
\end{figure}

We consider the scattering of water waves by a rectangular anisotropic elastic plate situated at the surface of the water and having negligible draft\change{, meaning that its draft is assumed to be much smaller than the plate's in-plane dimensions and the wavelength}. We adopt a three-dimensional Cartesian coordinate system $(x,y,z)$ for the fluid domain, where $x$ and $y$ are horizontal coordinates parallel to the plate edges, and the $z$ coordinate points vertically upward (i.e., it is normal to the plate at rest). At equilibrium, the fluid occupies the region $\Omega = \R^2\times(-H,0)$, where $H$ is the depth of the fluid, and the plate occupies the region $\change{\Gamma}\times \{0\}$, where $\change{\Gamma}=(0,a)\times(0,b)$. A schematic of the geometry is shown in Figure \ref{fig:schematic}.

The fluid is assumed to be \change{incompressible, inviscid and undergoing irrotational flow}, meaning that the time dependent velocity potential $\Phi$ satisfies
\begin{equation}
\bigtriangleup\Phi=0\quad\forall(x,y,z)\in\Omega,
\end{equation}
where $\Phi=\Phi(x,y,z,t)$ and $\bigtriangleup=\partial_x^2+\partial_y^2+\partial_z^2$. The velocity of the fluid is given by $\mathbf{v}=\nabla\Phi$.  There is no flow normal to the seabed at $z=-H$, which implies
\begin{equation}
    \partial_z\Phi=0\quad z=-H, (x,y)\in\R^2.
\end{equation}
The free surface of the fluid occupies the region $(x,y)\in\mathbb{R}^2\setminus \change{\Gamma}$. The free surface of the fluid is governed by the kinematic and dynamic boundary conditions, which are
\begin{subequations}\label{fs_eqs}
\begin{align}
\partial_t\zeta &= \partial_z\Phi{\biggr\rvert}_{z=0},\\
\zeta&=-\frac{1}{g}\partial_t\Phi{\biggr\rvert}_{z=0},
\end{align}
\end{subequations}
respectively, where $\zeta=\zeta(x,y,t)$ is the displacement of the free surface of the fluid from equilibrium and $g$ is the acceleration due to gravity.

The plate, which has uniform thickness $h$, is assumed to be thin, which means $a,b\gg h$. The motion of the plate is modelled using \change{Kirchhoff}--Love theory. \ctwo{We assume that in-plane deformations can be neglected because these will be negligible compared to bending. The Lagrangian of the plate is
\begin{equation}
\mathcal{L}\{W\}=T\{W\}-U_{\rm{strain}}\{W\}-U_{\rm{pressure}}\{W\}
\end{equation}
where $W(x,y,t)$ is the vertical displacement of the plate. The kinetic energy, strain energy due to bending and potential energy due to pressure are
\begin{subequations}
\begin{align}
T\{W\}&=\frac{1}{2}\rho h\iint_\Gamma (\partial_t W)^2\upd x\,\upd y\\
U_{\rm{strain}}\{W\}&=\frac{1}{2}\iint_\Gamma \mathcal{B}(W,W)\upd x\,\upd y\\
U_{\rm{pressure}}\{W\}&=-\iint_\Gamma PW\upd x\,\upd y,
\end{align}
\end{subequations}}
respectively, $\rho$ is the uniform density of the plate and $P(x,y,t)$ is the fluid pressure under the plate \citep{grossi2003,Reddy_2006,grossi2008}. The symmetric bilinear operator $\mathcal{B}$ is given by
\begin{equation}
\mathcal{B}(W_1,W_2)=\begin{bmatrix}
    \partial_x^2 W_1&\partial_y^2  W_1&\partial_x\partial_y W_1
\end{bmatrix}\begin{bmatrix}
    D_{11}&D_{12}&2D_{16}\\
    D_{12}&D_{22}&2D_{26}\\
    2D_{16}&2D_{26}&4D_{66}    
\end{bmatrix}\begin{bmatrix}
    \partial_x^2W_2\\\partial_y^2 W_2\\\partial_x\partial_y W_2
\end{bmatrix},
\end{equation}
where $D_{ij}$ are the rigidity coefficients which describe the material's resistance to bending and twisting \citep{Reddy_2006}. \ctwo{The action over an arbitrary time interval $(t_0,t_1)$ and its first variation are
\begin{align}
    \mathcal{A} &=\int_{t_0}^{t_1}\mathcal{L}\{W\}\upd t\\
    \delta\mathcal{A}&=\int_{t_0}^{t_1}\delta\mathcal{L}\{\delta W,W\}\upd t\\
    &=\int_{t_0}^{t_1}\int_\Gamma\left\{\rho h\,\delta W\, \partial_t^2 W+\mathcal{B}(\delta W,W)-P\delta W \right\}\upd x\,\upd y\,\upd t,
\end{align}
respectively, where $\delta W$ is an admissible variation of $W$. The principle of least action implies that $\delta \mathcal{A}=0$ for arbitrary admissible variations $\delta W$ over arbitrary time intervals $(t_0,t_1)$, which yields
\begin{equation}\label{variational_form_dynamic_plate}
\iint_\Gamma\left\{\rho h\,\delta W\,\partial_t^2 W+\mathcal{B}(\delta W,W)-P\delta W\right\}\,\upd x\,\upd y=0.
\end{equation}}
The strong form of \eqref{variational_form_dynamic_plate} is
\begin{equation}\label{transient_problem}
\rho h\partial_t^2{W}+\mathcal{D}W=P,
\end{equation}
where the fourth order differential operator $\mathcal{D}$ is given by
\begin{equation}
    \mathcal{D}=D_{11}\partial_{x}^4+4D_{16}\partial_{x}^3\partial_{y}+2(D_{12}+2D_{66})\partial_{x}^2\partial_{y}^2+4D_{26}\partial_{x}\partial_{y}^3+D_{22}\partial_{y}^4.
\end{equation}
\ctwo{Note that in the isotropic case, we have
\begin{subequations}
\begin{align}
D_{11}=D_{22}&=D,\\
D_{12}&=\nu D,\\
D_{66}&=\tfrac{1}{2}(1-\nu)D,\\
D_{16}=D_{26}&=0,\\
\text{where }D&=\frac{Eh^3}{12(1-\nu^2)},
\end{align}
\end{subequations}
in which $E$ is Poisson's ratio and $\nu$ is the Young's modulus. In this case, the operator $\mathcal{D}$ is proportional to the biharmonic operator.}

\change{From the linearised Bernoulli equation, the fluid pressure is the sum of  hydrostatic and hydrodynamic pressure, namely $P=P_{\mathrm{hydrostatic}}+P_{\mathrm{hydrodynamic}}$, where
\begin{equation}
    P_{\mathrm{hydrostatic}}=-\rho_wgz\quad\text{and}\quad P_{\mathrm{hydrodynamic}}=-\rho_w \partial_t\Phi,
\end{equation}
respectively, in which $\rho_w$ is the fluid density. By assuming small motions, we linearise the hydrodynamic pressure about the equilibrium position of the plate at $z=0$, which leads to the following dynamic condition on the plate:
\begin{equation}
\rho h \partial_t^2{W}+\mathcal{D}W=-\rho_wgW-\rho_w\partial_t\Phi{\biggr\rvert}_{z=0}.
\end{equation}
The coupling between the plate and the fluid is completed using the following linearised kinematic condition
\begin{equation}
\partial_z\Phi{\biggr\rvert}_{z=0}=\partial_t{W},
\end{equation}
for $(x,y)\in \change{\Gamma}$. By equating the fluid velocity at the fluid-plate interface with the plate velocity, this condition assumes that the plate remains fully in contact with the fluid at all times. We note that linear hydroelasticity is well validated experimentally for sufficiently low ratios of amplitude to wavelength \citep{Montiel2013b,Montiel2013a}.}

In this paper, the plate edges are assumed to be either clamped, implying
\begin{subequations}\label{clamped_BCs}
    \begin{align}
    W=\partial_x W &= 0&x=0\text{ or }x=a,\\
    W=\partial_y W&=0&y=0\text{ or }y=b,
\end{align}
\end{subequations}
or free, implying
\begin{subequations}\label{Free_BCs}
\begin{align}
D_{11}\partial_x^2 W +2D_{16}\partial_x\partial_y W+D_{12}\partial_y^2 W&=0&x=0\text{ or }x=a,\\
D_{11}\partial_x^3 W+4 D_{16}\partial_x^2\partial_y W+(D_{12}+4D_{66})\partial_x\partial_y^2W+2D_{26}\partial_y^3 W&=0&x=0\text{ or }x=a,\\
D_{22}\partial_y^2 W+2D_{26}\partial_x\partial_y W+D_{12}\partial_x^2 W&=0&y=0\text{ or }y=b,\\
D_{22}\partial_y^3 W+4D_{26}\partial_y^2\partial_x W+(D_{12}+4D_{66})\partial_y\partial_x^2W+2D_{16}\partial_y^3 W&=0&y=0\text{ or }y=b,\\
D_{16}\partial_x^2W+D_{26}\partial_y^2W+2D_{66}\partial_{x}\partial_yW&=0&(x,y)\in\Lambda,
\end{align}
\end{subequations}
where $\Lambda=\{(0,0),(a,0),(0,b),(a,b)\}$ are the corners of the plate \citep{grossi2003}, \ctwo{or simply supported, implying \cite[\textsection 6.1.2 of][]{Reddy_2006}
\begin{subequations}\label{SS_bcs}
\begin{align}
    W=D_{11}\partial_x^2 W+D_{12}\partial_y^2 W &= 0&x=0\text{ or }x=a,\\
    W=D_{12}\partial_x^2 W+D_{22}\partial_y^2 W&=0&y=0\text{ or }y=b.
\end{align}
\end{subequations}}
We note that while our framework can be extended to other boundary condition types (e.g. moored edges) or to have different boundary conditions on different edges, we do not explore this here.

\subsection{Time harmonic formulation}
We assume a time harmonic motion with angular frequency $\omega$, which implies that we can write
\begin{subequations}
\begin{align}
\Phi(x,y,z,t)&=\Re\{\phi(x,y,z)\e^{-\upi\omega t}\}&(x,y,z)\in\Omega,\\
\zeta(x,y,t)&=\Re\{\eta(x,y)\e^{-\upi\omega t}\}&(x,y)\in\R^2\setminus \change{\Gamma},\\
W(x,y,t)&=\Re\{\w(x,y)\e^{-\upi\omega t}\}&(x,y)\in \change{\Gamma}.
\end{align}
\end{subequations}
This leads to the following boundary value problem for $\phi$:
\begin{subequations}\label{coupled_bvp}
\begin{align}
\bigtriangleup\phi&=0&(x,y,z)\in\Omega,\label{laplace_fd}\\
\partial_z\phi&=0&z=-H,\quad(x,y)\in\mathbb{R}^2,\label{seabed_fd}\\
\partial_z\phi&=\alpha\phi&z=0,\quad(x,y)\in\mathbb{R}^2\setminus \change{\Gamma},\label{fs_fd}\\
\partial_z\phi&=-\upi\omega \w&z=0,\quad(x,y)\in \change{\Gamma},\label{plate_kinematic_fd}\\
\sqrt{r}(\partial_r-\upi k)(\phi-\phi^{\mathrm{inc}})&\to 0&\text{as }r\to\infty,\label{plate_sommerfeld}
\end{align}
with $\alpha = \omega^2/g$, which must be solved in tandem with the plate equation
\begin{equation}\label{coupled_plate_dynamic_fd}
    \change{\mathcal{D}}\w-\omega^2\rho h\w+\rho_w g\w=\upi\omega\rho_w\phi{\biggr\rvert}_{z=0},
\end{equation}
\end{subequations}
for $(x,y)\in \change{\Gamma}$, where the plate satisfies \ctwo{either clamped \eqref{clamped_BCs}, free \eqref{Free_BCs} or simply-supported \eqref{SS_bcs} boundary conditions, as prescribed.} \change{In the Sommerfeld condition \eqref{plate_sommerfeld}, $r=\sqrt{x^2+y^2}$, $k$ is the positive real root of the dispersion relation
\begin{equation}\label{dispersion_relation}
k\tanh(k H)=\alpha,
\end{equation}
and $\phi^{\rm inc}$ is the prescribed incident wave.}
\section{Diffraction by a rigid plate}\label{rigid_plate_sec}
We first consider the diffraction potential $\phi^{\mathrm{di}}$, which describes the scattering of a prescribed incident wave $\change{\phi^{\rm inc}}$ by a plate held fixed in a flat configuration at the fluid surface. This potential satisfies the boundary value problem consisting of \eqref{laplace_fd}, \eqref{seabed_fd}, \eqref{fs_fd}, \eqref{plate_sommerfeld}, together with the no flow condition on the plate
\begin{equation}\label{diffraction_BC}
\partial_z[\change{\phi^{\rm inc}}+\phi^{\mathrm{di}}]=0\quad\text{for } z=0,\quad (x,y)\in \change{\Gamma}.
\end{equation}
The unknown diffraction potential is obtained by solving a boundary integral equation.

\subsection{Green's function}\label{greens_fn_subsec}
The Green's function for a three-dimensional fluid of constant depth \change{with a point source on the surface of the fluid} satisfies
\begin{subequations}\label{Greens_fn_pde}
\begin{align}
    \bigtriangleup G &=\change{0}&(x,y,z)\in\Omega,\\
    \change{\partial_z G- \alpha G} &\change{=-\delta (x)\delta(y)}&z=0,\label{Green_FD}\\
    \partial_zG&=0&\change{z=-H},\\
    \sqrt{r}(\partial_rG-\upi kG)&\to0&\text{as }\change{r}\to\infty,
\end{align}
\end{subequations}
where $\delta(x)$ is the Dirac delta. \change{For $z=0$}, the Green's function has the following representation 
\begin{equation}\label{greens1}
-4\pi G(r)=\frac{1}{r}+\frac{1}{\sqrt{r^2+4H^2}}+\Uint_0^\infty\frac{2(\mu+\alpha)\e^{-\mu H}\cosh^2\mu H}{\mu\sinh\mu H-\alpha\cosh\mu H} \mathrm{J}_0(\mu r)\upd\mu,
\end{equation}
where $\mathrm{J}_0$ is the zeroth order Bessel function of the first kind and $\Uint$ denotes complex contour integration beneath any poles on the real line. \change{We note that \eqref{greens1} is equivalent to the submerged source Green's function given by \citet[Equation B.86]{Linton2001}, in the limit as the point source approaches the surface.} \change{A derivation of the Green's function using a Hankel transform is given in Appendix \ref{Greens_derivation}}. To evaluate the above integral numerically, we first write $\mathrm{J}_0(x)=\frac{1}{2}(\mathrm{H}_0^{(1)}(x)+\mathrm{H}_0^{(2)}(x))$ (where $\mathrm{H}_0^{(1)}$ and $\mathrm{H}_0^{(2)}$ are the zeroth order Hankel functions of the first and second kind, respectively) and deform the contour of integration using residue calculus. We obtain the following rapidly converging representation of the Green's function:
\begin{align}
-4\pi G(r)&=\frac{1}{r}+\frac{1}{\sqrt{r^2+4H^2}}+2\Re\left\{\int_C\frac{(\mu+\alpha)\e^{-\mu H}\cosh^2\mu H}{\mu\sinh\mu H-\alpha\cosh\mu H} \mathrm{H}_0^{(1)}(\mu r)\upd\mu\right\}\nonumber\\
&\qquad+\change{\frac{2\pi\upi}{H}\left(1+\frac{\sinh 2kH}{2kH}\right)^{-1}}\cosh^2(kH) \mathrm{H}_0^{(1)}(kr),\label{greens2}
\end{align}
where $C$ is parametrised by $\e^{\upi \pi/4}t$ for $0\leq t<\infty$. \change{Note that in order to consider infinite depth, we need only replace \eqref{greens1} with the corresponding infinite depth Green's function.} The integral in \eqref{greens2} is calculated using the MATLAB routine \verb|integral|.

\subsection{Integral equation for diffraction problem}\label{diffraction_BIE}
\change{We assume $\phi^{\mathrm{di}}$ can be written as a distribution of sources of the form
\begin{equation}
\phi^{\mathrm{di}}(\mathbf{x})=\iint_{\change{\Gamma}\times\{0\}}\sigma(\mathbf{x}^\prime)G(\mathbf{x},\mathbf{x}^\prime)\upd S_{\mathbf{x}^\prime},
\end{equation}
where, in abuse of notation, $G(\mathbf{x},\mathbf{x}^\prime)$ denotes the value of the Green's function at the point $\mathbf{x}=(x,y,z)$ for a source located at $\mathbf{x}^\prime=(x^\prime,y^\prime,z^\prime)$, whenever its arguments are bold. Then for $\mathbf{x}\in \change{\Gamma}\times\{0\}$ we compute
\begin{align}
(\alpha-\partial_z)\phi^{\mathrm{di}}(\mathbf{x})&=\iint_{\change{\Gamma}\times\{0\}}\sigma(\mathbf{x}^\prime)(\alpha-\partial_z)G(\mathbf{x},\mathbf{x}^\prime)\upd S_{\mathbf{x}^\prime}\nonumber\\
&=-\iint_{\change{\Gamma}\times\{0\}}\sigma(\mathbf{x}^\prime)\delta(x-x^\prime)\delta(y-y^\prime)\upd S_{\mathbf{x}^\prime}\nonumber\\
&=-\sigma(\mathbf{x}).
\end{align}
Thus
\begin{equation}\label{phi_di_representation}
\phi^{\mathrm{di}}(\mathbf{x})=\iint_{\change{\Gamma}\times\{0\}}(\alpha\phi^{\mathrm{di}}(\mathbf{x^\prime})-\partial_{z^\prime}\phi^{\mathrm{di}}(\mathbf{x^\prime}))G(\mathbf{x},\mathbf{x^\prime})\upd S_{\mathbf{x^\prime}}.
\end{equation}
An integral equation for $\phi^{\mathrm{di}}$ is obtained from the above by restricting to the plate and using \eqref{diffraction_BC}, namely
\begin{align}\label{BIE}
\phi^{\mathrm{di}}(\mathbf{x})=\alpha\iint_{\change{\Gamma}\times\{0\}}G(\mathbf{x}^\prime,\mathbf{x})(\phi^{\rm inc}(\mathbf{x}^\prime)+ \phi^{\mathrm{di}}(\mathbf{x}^\prime))\upd S_{\mathbf{x}^\prime}\quad\text{for all}\quad\mathbf{x}\in \change{\Gamma}\times\{0\}.
\end{align}
We note that the same integral equation may be obtained by using the Green's function for a submerged source \citep[as given by][Equation B.86]{Linton2001} and applying Green's third identity to $\phi^{\rm di}$. This involves a subtle point which arises when the source point in on the free surface and we replace the normal derivative of the Green function using the free-surface condition. This point is discussed in \cite{jgrfloecirc,meylan2002wave} where a similar Green's function method is used and also in \citet{buchner1993evaluation}.}

\subsection{Numerical solution using a constant panel method}\label{constant_panel}
The boundary integral equation above is solved using a constant panel method. To do so, we discretise the plate \change{surface $\change{\Gamma}\times\{0\}$} into a grid of $N_x\times N_y$ squares $\square_j$. Each square has dimensions $\Delta x\times\Delta x$ and midpoints $\mathbf{x}_j=(x_j,y_j,0)^\intercal$. Evaluated at nodes $\mathbf{x}_i$, equation \eqref{BIE} becomes
\begin{subequations}
\begin{align}
\phi^{\mathrm{di}}(\change{\mathbf{x}_i})&=\alpha\sum_{j=1}^{N_x\times N_y}\change{\iint_{\square_j}} G(\change{\mathbf{x}^\prime},\change{\mathbf{x}_i})\left(\change{\phi^{\rm inc}}(\change{\mathbf{x}^\prime})+\phi^{\mathrm{di}}(\change{\mathbf{x}^\prime})\right)\change{\upd S_{\mathbf{x}^\prime}}\\
&\approx \alpha\sum_{j=1}^{N_x\times N_y}\left(\change{\phi^{\rm inc}}(\change{\mathbf{x}_j})+\phi^{\mathrm{di}}(\change{\mathbf{x}_j})\right)\change{\iint_{\square_j}} G(\change{\mathbf{x}^\prime},\change{\mathbf{x}_i})\change{\upd S_{\mathbf{x}^\prime}},\label{constant_panel1}
\end{align}
\end{subequations}
where \eqref{constant_panel1} is the constant panel assumption, valid when $\change{\phi^{\rm inc}}$ and $\phi^{\mathrm{di}}$ are slowly varying relative to the grid spacing. Letting $(\boldsymbol{\phi}^{\mathrm{di}})_j=\phi^{\mathrm{di}}(x_j,y_j,0)$ and $(\boldsymbol{\phi}^{\mathrm{in}})_j=\change{\phi^{\rm inc}}(x_j,y_j,0)$, Equation \eqref{constant_panel1} may be recast in matrix form as
\begin{equation}
    (I-\alpha K)(\boldsymbol{\phi}^{\mathrm{in}}+\boldsymbol{\phi}^{\mathrm{di}})=\boldsymbol{\phi}^{\mathrm{in}},
\end{equation}
to be solved for $\boldsymbol{\phi}^{\mathrm{di}}$, where $I$ is the $(N_x\times N_y)$-dimensional identity matrix. The entries in the kernel matrix $K$ are chosen to approximate the integrals over the squares $\square_j$ appearing in \eqref{constant_panel1}, namely,
\begin{equation}
(K)_{ij}\approx\change{\iint_{\square_j} G(\mathbf{x}^\prime,\mathbf{x}_i)\upd S_{\mathbf{x}^\prime}}.
\end{equation}
We approximate the off diagonal entries of $K$ as
\begin{equation}
\change{(K)_{ij} = G(\mathbf{x_j},\mathbf{x}_i)\Delta x^2,}
\end{equation}
for $i\neq j$. This is equivalent to midpoint quadrature on the grid. \change{Greater care is applied to the integrals associated with the diagonal entries of $K$, namely}
\begin{subequations}
\begin{align}
\change{(K)_{ii}}&\change{=\iint_{\square_j} G(\mathbf{x}^\prime,\mathbf{x}_i)\upd S_{\mathbf{x}^\prime}}\\
&=\int_{-\Delta x/2}^{\Delta x/2}\int_{-\Delta x/2}^{\Delta x/2} G\left(\sqrt{x^2+y^2},0\right)\upd x\, \upd y,\label{greens_self_centered}
\end{align}
\end{subequations}
for all $i$ \change{(meaning this quantity only needs to be calculated once)}, where the second line follows from translation. The double integral is then calculated by (i) directly integrating the term $1/r$ in \eqref{greens2}, and (ii) using the MATLAB routine \verb|integral2| to \change{numerically} integrate the remaining part. \change{We remark that in preliminary tests of our method, the integral representation of the Green's function \eqref{greens2} converged faster than methods based on an equivalent series representation (i.e., one analogous to Equation B.91 in \citet{Linton2001}), presumably due to direct integration of the $1/r$ term that dominates in the near field}. Our numerical code for the diffraction problem was validated for the case of a circular plate, using a code developed to solve the problem of wave scattering by a partially submerged truncated circular cylinder, which is valid in the limit of negligible submergence \citep{wilks2025simultaneouslayoutdeviceparameter}. The aforementioned code uses separation of variables in cylindrical coordinates to solve the problem both beneath and exterior to the plate. Subsequently, continuity conditions between the interior and exterior regions reduce to integral equations that are solved numerically using a singularity respecting Galerkin method.

\section{\change{Dry m}odes of vibration of the plate}\label{vibration_modes_sec}
The modes of vibration of the plate are the eigenfunctions $\w_j$ of the operator $\mathcal{D}$, which satisfy
\begin{equation}\label{fd_problem}
        \mathcal{D}\w_j=\rho h\omega_j^2  \w_j,
\end{equation}
and the boundary conditions \ctwo{(either \eqref{clamped_BCs}, \eqref{Free_BCs} or \eqref{SS_bcs} for a clamped, free or simply-supported plate, respectively)}, where $\omega_j$ are the angular frequencies of the vibrational modes. \change{They are called dry modes because they solve the dynamic equation of the plate \eqref{transient_problem} in the absence of any forcing from the fluid (i.e. with $P=0$).} \change{The corresponding weak form is
\begin{equation}\label{weak_form}
\iint_{\change{\Gamma}} \mathcal{B}(\w_j,\delta \w)\upd x\, \upd y=\rho h\omega_j^2\iint_{\change{\Gamma}} \w\,\delta \w\,\upd x\, \upd y,
\end{equation}
where $\delta \w$ is an admissible variation of $\w_j$. We seek a numerical solution of \eqref{weak_form} using the Rayleigh--Ritz method, which is classical but included here for completeness \citep[see][for further discussion]{Reddy_2006}. To begin, we introduce a set of functions $\{\psi_1,\dots,\psi_{N_{RR}}\}$, which satisfy the boundary conditions \eqref{clamped_BCs}. We approximate $\ctwo{\w}_j$ in the form of a superposition over this set, namely
\begin{equation}\label{rayleigh_ritz_superposition}
    \w_j\approx a_1\psi_1+\dots+a_{N_{RR}}\psi_{N_{RR}},
\end{equation}
where $a_1,\dots,a_{N_{RR}}$ are unknown coefficients. Lastly, we substitute \eqref{rayleigh_ritz_superposition} into \eqref{weak_form} for all test functions $\psi_1,\dots,\psi_{N_{RR}}$, yielding the following system of equations for the unknown coefficients $a_\theta$:
\begin{equation}\label{RR_eig_matrix_problem}
\sum_{\theta=1}^{N_{RR}} a_\theta\iint_{\change{\Gamma}}\mathcal{B}(\psi_\theta,\psi_\tau)\upd x\,\upd y  =\rho h \omega_j^2\sum_{\theta=1}^{N_{RR}}a_\theta\iint_{\change{\Gamma}}\psi_\theta\psi_\tau\,\upd x\,\upd y.
\end{equation}}

\change{To proceed, a suitable set of functions must be selected for the Rayleigh--Ritz method. In the case that the plate is clamped, we construct such functions from the modes of vibration of a clamped-clamped Euler--Bernoulli beam of unit length, which, when normalised, satisfy
\begin{subequations}\label{beam_equation}
\begin{align}
\partial_x^4 u_m&=\kappa_m^4u_m,\\
u_m(0)&=u_m(1)=0,\\
 u_m^\prime(0)&=u_m^\prime(1)=0,\\
    \int_0^1 u_m(x)u_n(x)\upd x&=\delta_{mn},\label{beam_modes_orthonormality}
\end{align}
\end{subequations}
where $\kappa_m$ are the beam eigenvalues and $\delta_{mn}$ is the Kronecker delta.} Further details about these eigenmodes are given in Appendix \ref{beam_eigenfunction_sec}. \change{As considered by \citet{Young1950}, the basis for the Rayleigh--Ritz method is then taken to be
\begin{subequations}\label{abstract_RR_substitution}
\begin{align}
\psi_\theta(x,y)&=u_{m(\theta)}\left(\frac{x}{a}\right) u_{n(\theta)}\left(\frac{y}{b}\right),\\
a_\theta&=A_{m(\theta) n(\theta)}^{(j)},
\end{align}
\end{subequations}
where $\theta$ indexes all pairs $(m,n)\in\mathbb{N}^2$ such that $1\leq m\leq N_x$ and $1\leq n\leq N_y$, in which $N_x$ and $N_y$ are the number of beam modes used in the $x$ and $y$ dimension, respectively. Thus $N_{RR}=N_xN_y$. \change{We note that this choice of basis functions is advantageous since they are mutually orthogonal and satisfy the plate boundary conditions \eqref{clamped_BCs}, although other choices of basis functions could be made \citep[e.g. polynomials, see][Chapter 7]{Reddy_2006}}. Arguments of $\theta$ in indices $m$ and $n$ will be dropped in what follows.}

\change{With the substitutions given in \eqref{abstract_RR_substitution}, \eqref{RR_eig_matrix_problem} eventually becomes
\begin{align}\label{eigenvalue_problem}
&\sum_{m=1}^{N_x}\sum_{n=1}^{N_y}\left[\frac{D_{12}}{a^2b^2}\left( \mathcal{U}_{mp}^{[0,2]}\mathcal{U}_{nq}^{[2,0]}+\mathcal{U}_{mp}^{[2,0]}\mathcal{U}_{nq}^{[0,2]}\right)\right.+\frac{2D_{16}}{a^3b}\left(\mathcal{U}_{mp}^{[2,1]}\mathcal{U}_{nq}^{[0,1]}+\mathcal{U}_{mp}^{[1,2]}\mathcal{U}_{nq}^{[1,0]}\right)\nonumber\\&\qquad+\frac{2D_{26}}{ab^3}\left(\mathcal{U}_{mp}^{[0,1]}\mathcal{U}_{nq}^{[2,1]}+\mathcal{U}_{mp}^{[1,0]}\mathcal{U}_{nq}^{[1,2]}\right)+\frac{4D_{66}}{a^2b^2}\mathcal{U}_{mp}^{[1,1]}\mathcal{U}_{nq}^{[1,1]}\quad+\frac{D_{11}}{a^4}\mathcal{U}_{mp}^{[2,2]}\mathcal{U}_{nq}^{[0,0]}\nonumber\\&\left.\qquad+\frac{D_{22}}{b^4}\mathcal{U}_{mp}^{[0,0]}\mathcal{U}_{nq}^{[2,2]}\right]A_{mn}^{(j)}=\omega^2\rho h A_{pq}^{(j)},
\end{align}
for $1\leq p\leq N_x$ and $1\leq q\leq N_y$, where we have defined
\begin{equation}\label{U_inner_product}
    \mathcal{U}_{mp}^{[j,l]} = \int_0^1 u_m^{(j)}(x)u_p^{(l)}(x)\mathrm{d}x,
\end{equation}
in which $u_m^{(l)}$ denotes the $l$th derivative of $u_m$.} Equation \eqref{eigenvalue_problem} is an eigenvalue problem of dimension $N_xN_y$ for the coefficients $A^{(j)}_{mn}$, which we solve numerically. The eigenmodes $\w_j(x,y)$, which are recovered from \change{the coefficients $A^{(j)}_{mn}$,} are normalised so that
\begin{equation}\label{plate_modes_orthonormality}
\iint_{\Gamma} \ctwo{\w}_m(x,y)\ctwo{\w}_n(x,y)^*\,\upd x\upd y= \delta_{mn}.
\end{equation}

\change{In the case that the edges of the plate are free, we construct instead the functions used in the Rayleigh--Ritz method from the modes of vibration of a free-free Euler--Bernoulli beam of unit length, which, when normalised, satisfy
\begin{subequations}\label{beam_equation_free}
\begin{align}
\partial_x^4 u_m&=\kappa_m^4u_m,\\
u_m^{(2)}(0)&=u_m^{(2)}(1)=0,\\
 u_m^{(3)}(0)&=u_m^{(3)}(1)=0,\\
 \int_0^1 u_m(x)u_n(x)\upd x&=\delta_{mn}.
\end{align}
\end{subequations}
Use of these functions instead of the corresponding modes of a clamped-clamped beam yields the modes of the free-edge plate, because free edges are the natural boundary conditions.}

\begin{figure}
    \centering
    \includegraphics[width=\linewidth]{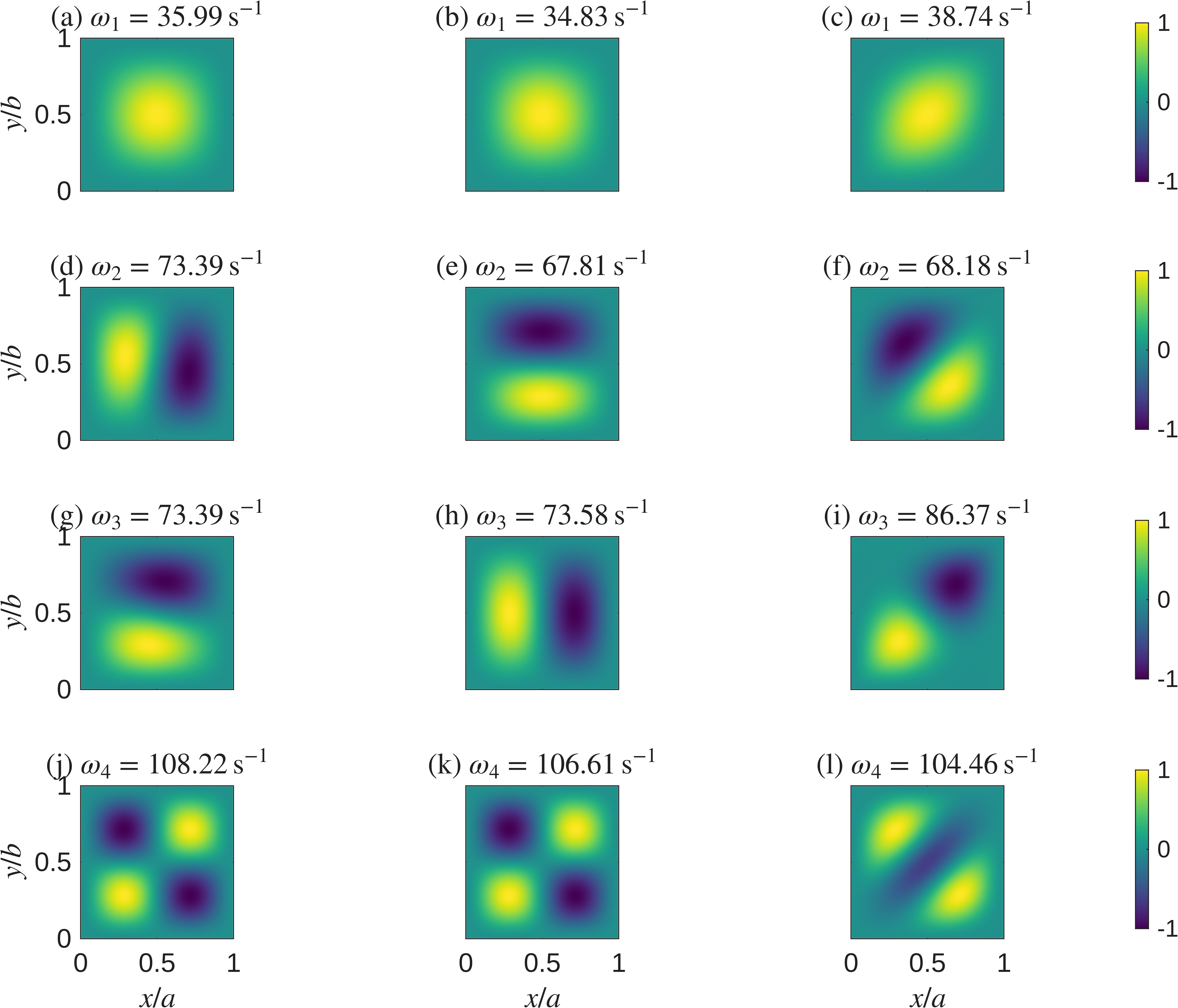}
    \caption{\change{First four mode shapes of a square (a,d,g,j) isotropic plate, (b,e,h,k) orthotropic plate and (c,f,i,l) anisotropic plate.} The rigidity coefficients associated with these terms are given in Table \ref{tab:rigidity_coeffs}. The remaining parameters are $a=b=1$\,m and $\rho h = 1$\,kg\,m$^{-1}$. Note the mode shapes have been rescaled as $\w_j(x,y)/p_j$, with $p_j$ chosen for presentation.}
    \label{fig:mode_shapes}
\end{figure}

\ctwo{Finally, in the case that the edges of the plate are simply supported, we construct the functions used in the Rayleigh--Ritz method from the modes of vibration of a simply supported-simply supported Euler--Bernoulli beam of unit length, which, when normalised, satisfy
\begin{subequations}\label{beam_equation_SS}
\begin{align}
\partial_x^4 u_m&=\kappa_m^4u_m,\\
u_m(0)&=u_m(1)=0,\\
 u_m^{(2)}(0)&=u_m^{(2)}(1)=0,\\
 \int_0^1 u_m(x)u_n(x)\upd x&=\delta_{mn}.
\end{align}
\end{subequations}}
We take $N_x/a=N_y/b = 20$ throughout this paper. We have validated our method for computing the \change{dry} modes using results provided by \change{\citet[][Table C1]{LEISSA1973257}} in the case where the plate is isotropic, and results provided in \citet{10.1115/1.4053090} in the case where the plate is anisotropic, with excellent agreement---see the first and third \change{columns} of Figure \ref{fig:mode_shapes}.

\begin{table}
\centering
\begin{tabular}{l|lll}
         & Isotropic &Orthotropic& Anisotropic \\ \hline
$D_{11}$ & 1 &1        & 1           \\
$D_{22}$ & 1   &\change{0.75}      & 1           \\
$D_{12}$ & 0.3   &\change{0.225}      & 0.9082      \\
$D_{16}$ & 0   &0      & 0.6724      \\
$D_{26}$ & 0   &0      & 0.6724      \\
$D_{66}$ & 0.35    &\change{0.4663}     & 0.9341     
\end{tabular}
\caption{Rigidity coeffficients used to generate the results in this paper. \change{The Poisson's ratio associated with the isotropic plate rigidity coefficients is $0.3$.} \change{The rigidity coefficients of the orthotropic plate correspond to a material under plane-stress conditions for which $E_y/E_x = 3/4$, $G_{xy}/E_x=1/2$ (where $E_x$ and $E_y$ are the Young's moduli in the $x$ and $y$ directions, respectively, and $G_{xy}$ is the shear modulus) and the Poisson's ratio is $0.3$.} The coefficients for the anisotropic plate are as in \cite{10.1115/1.4053090}. All units are in ${\rm Pa}\,{\rm m}^{\change{3}}$. }
\label{tab:rigidity_coeffs}
\end{table}

\section{Radiation potentials}\label{radiation_potentials_sec}
For each vibration mode of the plate $\w_j$, we compute the radiation potential $\phi^{\mathrm{ra}}_j$, namely the velocity potential of the fluid induced by the plate motion in the form $\w=\Re\{\w_j(x,y)\e^{-\upi\omega t}\}$. The radiation potential $\phi^{\mathrm{ra}}_j$ satisfies the boundary value problem consisting of \eqref{laplace_fd}, \eqref{seabed_fd}, \eqref{fs_fd}, \eqref{plate_sommerfeld} \change{in the case $\phi^{\mathrm{inc}}=0$}, and the kinematic condition
\begin{equation}\label{radiation_kinematic}
\partial_z\phi^{\mathrm{ra}}_j=-\upi\omega \w_j\quad\change{\text{for }(x,y)\in\Gamma,\quad z=0}.
\end{equation}
Following the steps outlined in \textsection\ref{diffraction_BIE}, \change{we use the Green's function (introduced in \textsection\ref{greens_fn_subsec}) to derive an integral equation for the radiation potential. We obtain}
\begin{equation}\label{phi_ra_representation}
\phi^{\mathrm{ra}}_j(\mathbf{x}) =  \change{\iint_{\Gamma\times\{0\}}}G(\mathbf{x}^\prime,\mathbf{x})(\alpha \phi^{\mathrm{ra}}_j(\mathbf{x}^\prime)+\upi\omega \w_j(\mathbf{x}^\prime))\upd S_{\mathbf{x}^\prime},
\end{equation}
where, in abuse of notation, we have written $\ctwo{\w}_j(\mathbf{x^\prime})=\ctwo{\w}_j([x^\prime,y^\prime,0]^\intercal)=\ctwo{\w}_j(x^\prime,y^\prime)$. Applying the constant panel discretisation method used in \eqref{constant_panel} then yields the matrix equation
\begin{equation}
    (I-\alpha K)\left(\boldsymbol{\phi}^{\mathrm{ra}}_j+\frac{\upi g}{\omega}\mathbf{w}_j\right) = \frac{\upi g}{\omega}\mathbf{w}_j,
\end{equation}
to be solved for $\boldsymbol{\phi}^{\mathrm{ra}}_j$, where $(\boldsymbol{\phi}^{\mathrm{ra}}_j)_i=\phi^{\mathrm{ra}}_j(x_i,y_i)$ and $(\mathbf{w}_j)_i = \w_j(x_i,y_i)$.

\section{Scattering by an elastic plate}\label{coupled_sec}
\noindent The solution to the coupled fluid-plate system of equations \eqref{coupled_bvp} is obtained by expanding in the \change{dry modes of vibration of the plate, following the theory of \citet{newman1994wave}.} That is, the total potential is taken to be a superposition of the incident potential, the diffraction potential plus a weighted sum of the radiation potentials
\begin{equation}\label{radiation_expansion}
    \phi = \change{\phi^{\rm inc}}+\phi^{\mathrm{di}}+\sum_{j=1}^\infty c_j \phi^{\mathrm{ra}}_j,
\end{equation}
where the coefficients $c_j$ are unknown. By the kinematic condition \eqref{plate_kinematic_fd}, the deformation of the plate is of the form
\begin{equation}
    \w = \sum_{j=1}^\infty c_j \w_j.
\end{equation}
The dynamic condition \eqref{coupled_plate_dynamic_fd}, when multiplied by $\w_m^*$ and integrated over the plate, then yields
\begin{equation}
\rho h\omega_m^2c_m-\rho h\omega^2c_m+\rho_wgc_m = \upi\omega\rho_w\change{\iint_{\Gamma\times\{0\}} }\phi \w_m^*\upd S,
\end{equation}
where we have used \eqref{fd_problem}. Substituting \eqref{radiation_expansion} into the above then yields
\begin{equation}\label{mode_coefficient_system}
\rho h\omega_m^2 c_m -\rho h\omega^2 c_m +\rho_wg c_m- \upi\omega \rho_w\sum_{j=1}^\infty \left(\change{\iint_{\Gamma\times\{0\}}}\phi^{\mathrm{ra}}_j \w_m^*\upd S\right)c_j = \upi\omega\rho_w\change{\iint_{\Gamma\times\{0\}}}(\change{\phi^{\rm inc}}+\phi^{\mathrm{di}})\w_m^*\upd S.
\end{equation}
Numerical evaluation of the integrals in the above, and truncation of the series after $N$ terms, yields an $N\times N$ system for the unknown coefficients $c_1,\dots,c_N$. Equation \eqref{mode_coefficient_system} becomes
\begin{equation}
    (\change{K_{\mathrm{stiff}}}-\omega^2M+C-\omega^2A-\upi\omega B)\mathbf{c}=\mathbf{f},
\end{equation}
where $\change{K_{\mathrm{stiff}}}$, $M$ and $C$ are the stiffness, mass and hydrostatic restoration matrices of the system, respectively, given by
\begin{subequations}
    \begin{align}
    \change{K_{\mathrm{stiff}}} &= \rho h\ \mathrm{diag}(\omega_m^2),\\
    M &= -\rho h I,\\
    C &= \rho_w g I,
\end{align}
where $I$ is the $N\times N$ identity matrix. Moreover, the real-valued added mass and damping matrices $A$ and $B$, respectively, are such that
\begin{equation}
    -\omega^2A_{mj}-\upi\omega B_{mj}=-\upi\omega \rho_w\change{\iint_{\Gamma\times\{0\}}}\phi^{\mathrm{ra}}_j \w_m^*\upd S,
\end{equation}
the vector of unknown coefficients is $\mathbf{c}=[c_1,\dots,c_N]^\intercal$, and the forcing vector is
\begin{equation}
(\mathbf{f})_m=\upi\omega\rho_w\change{\iint_{\Gamma\times\{0\}}}(\change{\phi^{\rm inc}}+\phi^{\mathrm{di}})\w_m^*\upd S.
\end{equation}
\end{subequations}
Note that in this paper $N=30$ plate modes were considered, which was considered sufficient based on the energy balance validation in the following section. \change{Also note that our implementation of this method has been validated using code developed by \citet{meylan2002wave} for the case of a square isotropic plate with free edges.}

\section{Energy balance calculations}\label{energy_balance_sec}
To verify our calculations, an energy balance identity is derived for the case in which the plate is excited by an incident plane wave of amplitude $A$ and incident angle $\chi$, namely
\begin{equation}\label{incident_plane_wave}
\change{\phi^{\rm inc}}(x,y,z)=\frac{-\upi gA}{\omega}\frac{\cosh(k(z+H))}{\cosh(kH)}\e^{\upi k(x\cos\chi+y\sin\chi)},
\end{equation}
such that the corresponding free surface elevation is $\eta^{\rm in}(x,y)=\e^{\upi kx}$. Let $(r,\theta,z)$ be the system of cylindrical coordinates with $x=r\cos\theta$ and $y=r\sin\theta$. For large $r$, the potential can be approximated as \citep{Mei2005}
\begin{equation}\label{far_field_mei}
\phi\sim\frac{-\upi gA}{\omega}\frac{\cosh(k(z+H))}{\cosh(kH)}\left(\e^{\upi kr\cos(\theta-\chi)}+\sqrt{\frac{2}{\pi kr}}f(\theta)\e^{\upi(k r-\pi/4)}\right).
\end{equation}
In the above, the first term in the brackets describes the incident wave and the second term describes the scattered wave. Thus, the far field pattern $f$ must be derived from the scattered field. We first note that the total scattered field is of the form
\begin{equation}\label{scattered_field_general_form}
\phi-\change{\phi^{\rm inc}}=\change{\iint_{\Gamma\times\{0\}}}G(\mathbf{x}^\prime,\mathbf{x})u(\mathbf{x}^\prime)\upd S_{\mathbf{x}^\prime},
\end{equation}
where, with reference to \eqref{phi_di_representation} and \eqref{phi_ra_representation},
\begin{equation}
u(x,y)=\alpha(\change{\phi^{\rm inc}}(x,y,0)+\phi^{\mathrm{di}}(x,y,0))+\sum_{j=1}^\infty c_j(\alpha\phi_j^{\mathrm{ra}}(x,y,0)+\upi\omega\w_j(x,y)).
\end{equation}
Asymptotically, the Green's function is
\begin{equation}
G(r,\theta,z;0,0)\sim-\frac{\upi}{2}\frac{k\cosh(k(z+H))}{k H\sech(kH)+\sinh(kH)}\sqrt{\frac{2}{\pi kr}}\e^{\upi(kr-\pi/4)}.
\end{equation}
Suppose the source point of the Green's function is shifted to $(x^\prime,y^\prime)$. Let $d(x^\prime,y^\prime)$ and $\alpha(x^\prime,y^\prime)$, respectively, be the radial and angular coordinates of the point $(x^\prime,y^\prime)$ with respect to the global origin $(0,0)$. Then as $r\to\infty$, the Green's function is asymptotically given by \citep{Falnes2002}
\begin{align}
G(r,\theta,z;x^\prime,y^\prime)&\sim-\frac{\upi}{2}\frac{k\cosh(k(z+H))}{k H\sech(kH)+\sinh(kH)}\sqrt{\frac{2}{\pi kr}}\nonumber\\
&\qquad\qquad\times\exp(\upi(kr-kd(x^\prime,y^\prime)\cos(\alpha(x^\prime,y^\prime)-\theta)-\pi/4)).
\end{align}
With reference to \eqref{far_field_mei}, substitution of the above expression into \eqref{scattered_field_general_form} eventually yields
\begin{equation}
f(\theta)=\frac{k\omega}{2g[kH\sech^2(kH)+\tanh(kH)]}\iint_{\change{\Gamma}}u(x^\prime,y^\prime)\e^{-\upi kd(x^\prime,y^\prime)\cos(\alpha(x^\prime,y^\prime)-\theta)}\upd y^\prime\upd x^\prime.
\end{equation}
Conservation of energy implies that the optical theorem holds, which is \citep{Mei2005}
\begin{equation}\label{energy_identity}
\frac{1}{2\pi}\int_{-\pi}^\pi |f(\theta)|^2\upd\theta=-\Re\{f(\chi)\}.
\end{equation}
We have used this identity to verify our computations. Numerical parameters are chosen in order to obtain five figure agreement between the left and right hand sides of \eqref{energy_identity}. This was obtained for $\Delta x=(1/120)$\,m.

\section{Results}\label{results_sec}
Throughout this results section, we take $H=20$\,m and $\rho h=1$\,kg\,m$^{-1}$. We consider three cases of the rigidity coefficients, as outlined in Table \ref{tab:rigidity_coeffs}. We take the incident wave to be an incident plane wave of the form \eqref{incident_plane_wave} with $A=1$\,m and $\chi=0$ (i.e., it is travelling in the positive $x$ direction) unless otherwise stated. The time averaged kinetic energy of the plate is
\begin{equation}
\bar{E}_{\mathrm{kin}}=\frac{\rho h \omega^2}{4}\int_0^a\int_0^b|\w(x,y)|^2\upd y \upd x =\frac{\rho h\omega^2}{4}\sum_{j=1}^\infty|c_j|^2.
\end{equation}
This quantity is used to identify resonant frequencies of the plate.

\begin{figure}
    \centering
    \includegraphics[width=\linewidth]{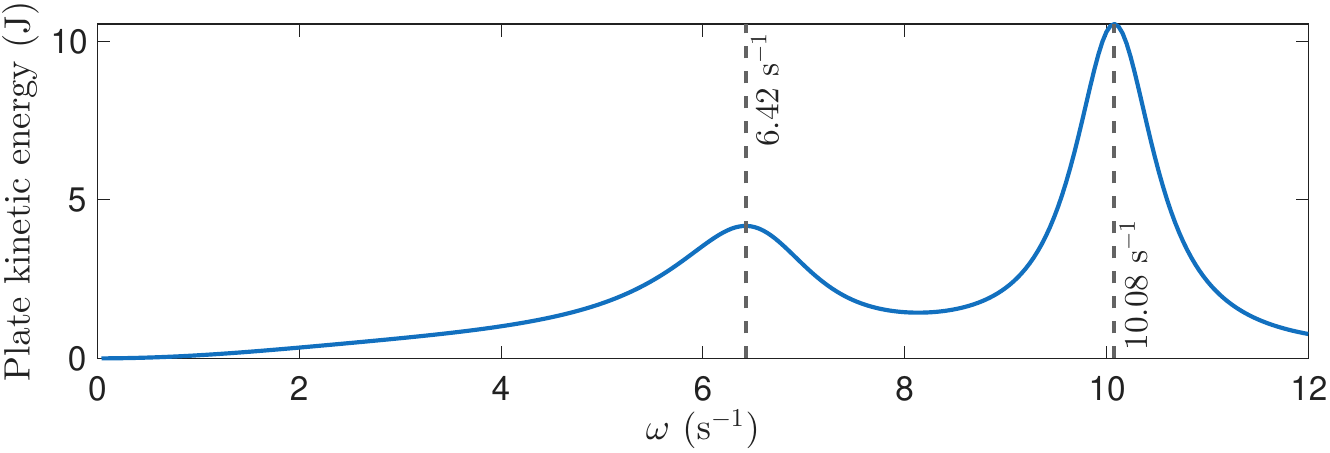}
    \caption{Kinetic energy spectrum for a square isotropic plate \change{with clamped edges}, with $a=b=1$\,m.}
    \label{fig:isotropic_spectrum}
\end{figure}

\begin{figure}
    \centering
    \includegraphics[width=\linewidth]{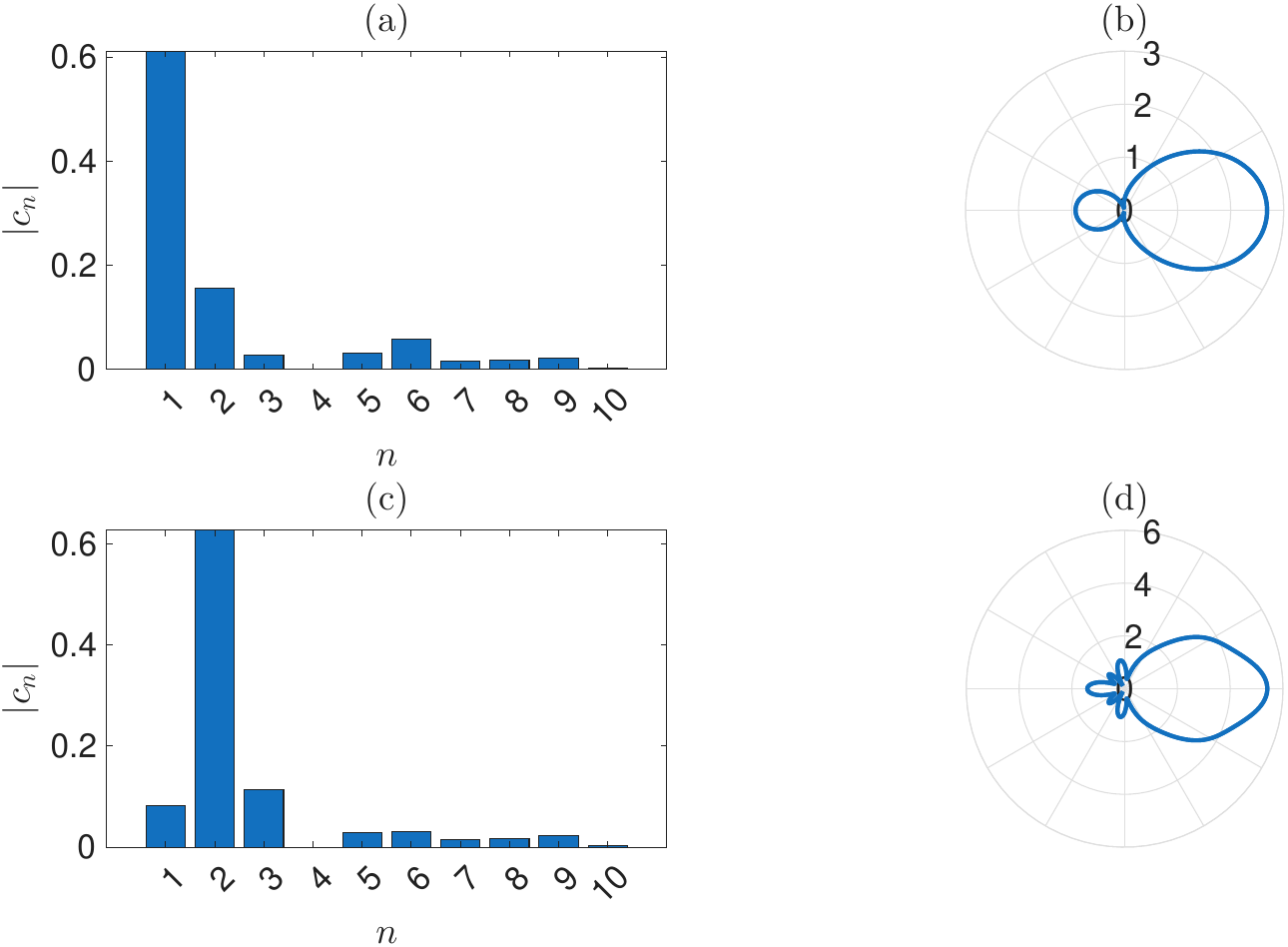}
    \caption{\change{Dry m}ode expansion coefficient magnitudes (left column) and far field patterns (i.e., polar plots of $f(\theta)$, right column) at resonant frequencies of the square isotropic plate \change{with clamped edges and} with $a=b=1$\,m. In particular, the resonant frequencies are (a,b) \change{$\omega=6.42$\,s$^{-1}$ and (c,d) $\omega=10.08$\,s$^{-1}$.}}
    \label{fig:isotropic_quantities}
\end{figure}

\begin{figure}
    \centering
    \includegraphics[width=\linewidth]{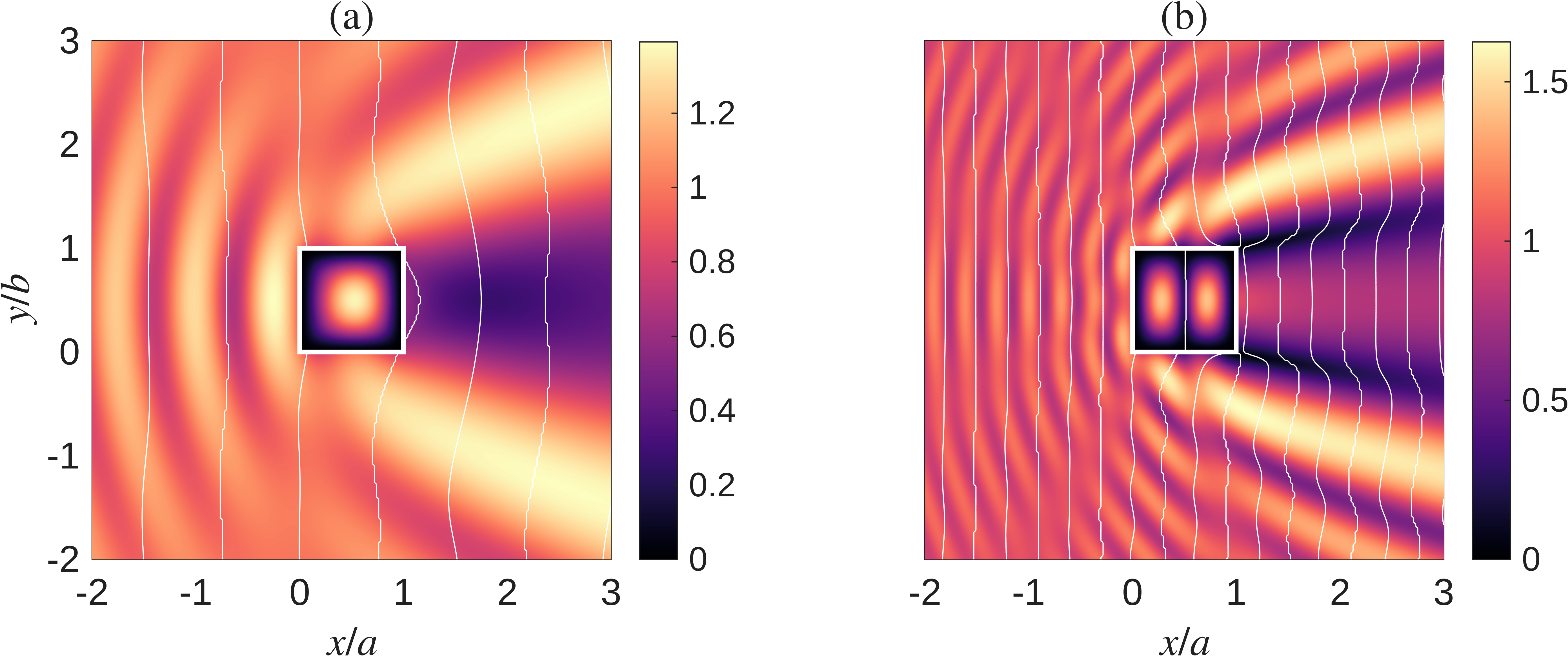}
    \caption{Surface elevation of the excited square isotropic plate \change{with clamped edges} ($a=b=1$\,m) at the resonant frequencies considered in Figure \ref{fig:isotropic_quantities}, namely \change{(a) $\omega=6.42$\,s$^{-1}$ and (b) $\omega=10.08$\,s$^{-1}$.} The surface elevation refers to $|w|$ over the plate, and $|\eta|$ otherwise. The absolute value of the surface elevation is indicated by the colour scale, with white isophasic contours indicating where the surface elevation is real. The boundary of the plate is marked with a thick white line.}
    \label{fig:isotropic_field}
\end{figure}

Figure \ref{fig:isotropic_spectrum} shows the kinetic energy spectrum for a square isotropic plate. The two peaks correspond to resonances of the plate, which we investigate in Figures \ref{fig:isotropic_quantities} and \ref{fig:isotropic_field}. We observe that the first resonance is predominantly associated with excitation of the first mode of vibration of the plate, whereas the second resonance is predominantly associated with excitation of the second and third modes in equal amounts. This is unsurprising because the second and third modes are degenerate, as shown in Figure \ref{fig:mode_shapes}. We note that in both cases, the fourth mode of the plate is not excited (i.e., $c_4=0$) because it is antisymmetric with respect to the line $y=b/2$, while the incident wave is symmetric with respect to this line. We remark that the far field patterns (Figure \ref{fig:isotropic_quantities}b and d) are symmetric with respect to $\theta=0$, and the surface elevations (Figure \ref{fig:isotropic_field}) are symmetric about $y=b/2$. 

\begin{figure}
    \centering
    \includegraphics[width=\linewidth]{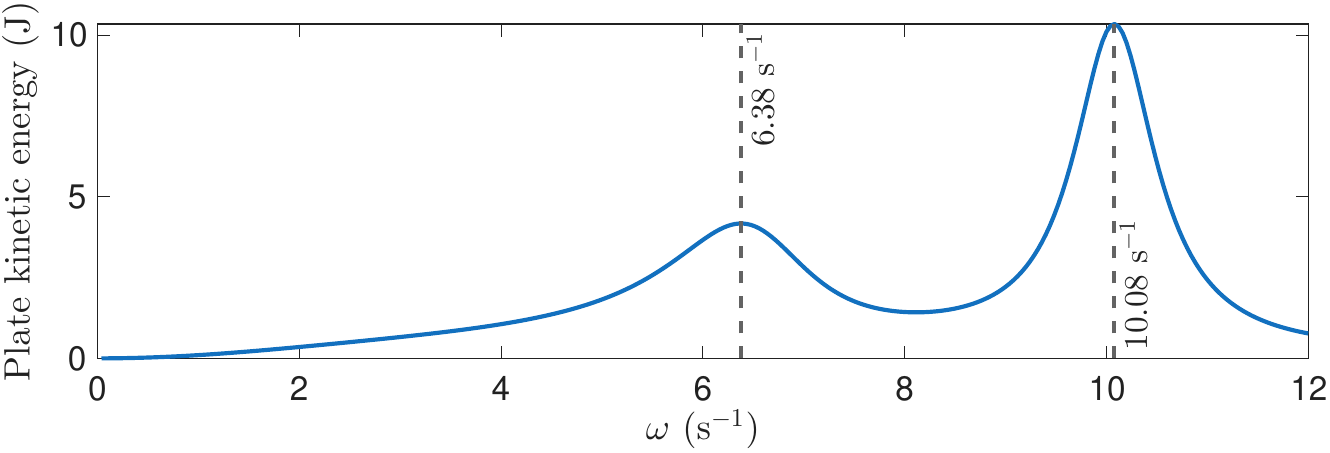}
    \caption{Kinetic energy spectrum for a square orthotropic plate \change{with clamped edges}, with $a=b=1$\,m.}
    \label{fig:orthotropic_spectrum}
\end{figure}

\begin{figure}
    \centering
    \includegraphics[width=\linewidth]{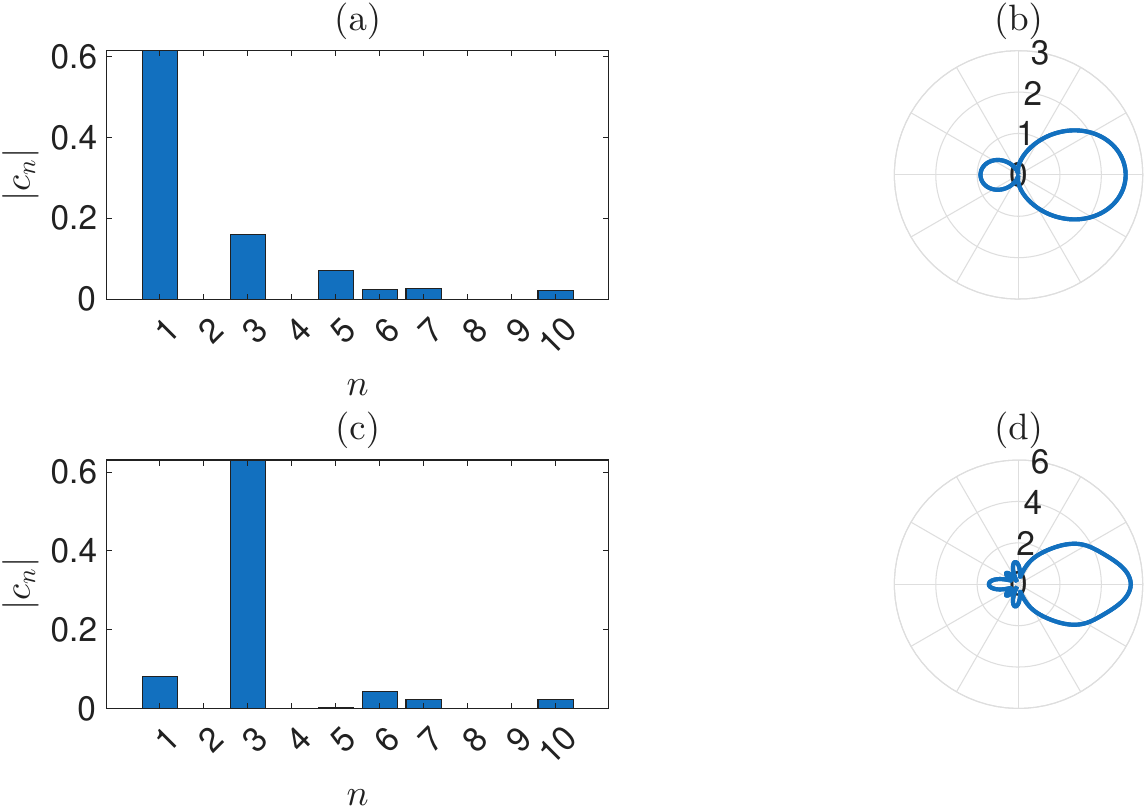}
    \caption{\change{Dry m}ode expansion coefficient magnitudes (left column) and far field patterns (right column) at resonant frequencies of the square orthotropic plate \change{with clamped edges and} with $a=b=1$\,m. The \change{resonant frequencies} are (a,b) \change{$\omega=6.38$\,s$^{-1}$ and (c,d) $\omega=10.08$\,s$^{-1}$.}}
    \label{fig:orthotropic_quantities}
\end{figure}

\begin{figure}
    \centering
    \includegraphics[width=\linewidth]{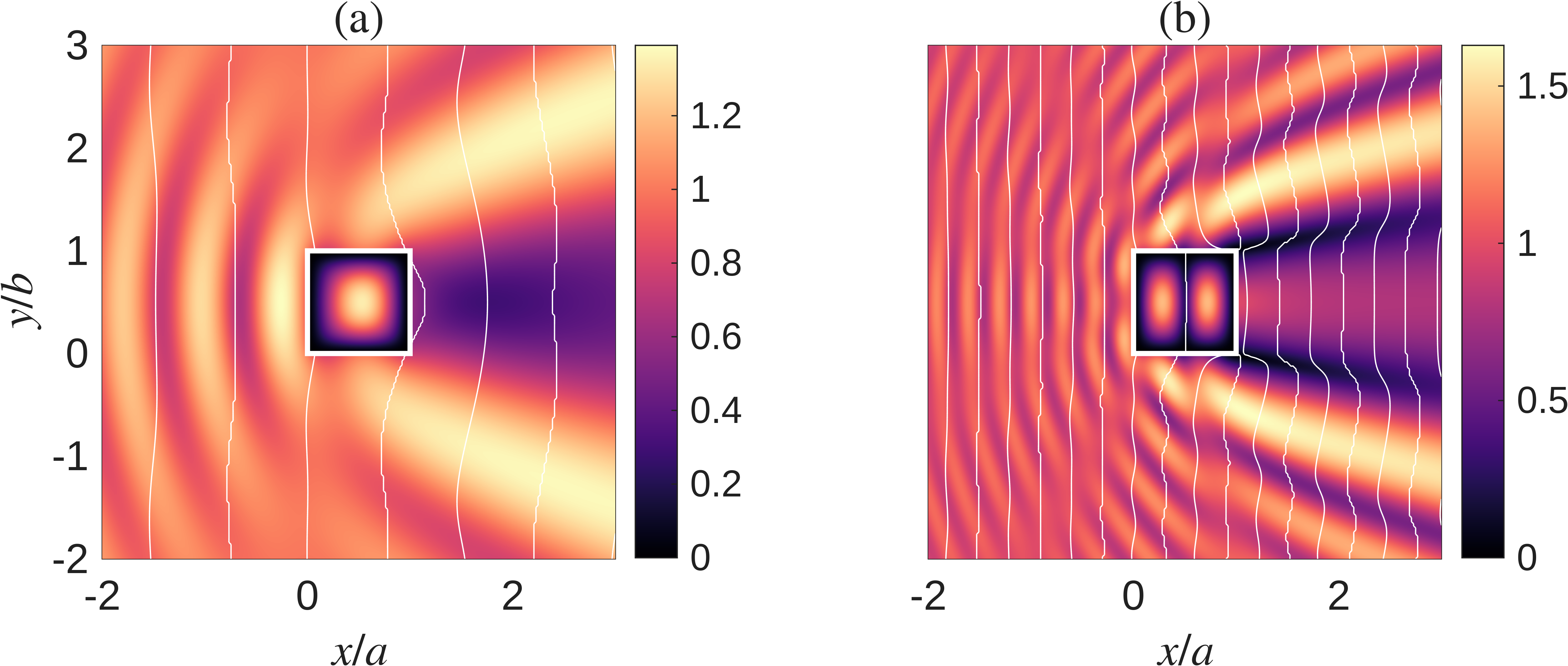}
    \caption{Surface elevation of the excited square orthotropic plate \change{with clamped edges} ($a=b=1$\,m) at the resonant frequencies considered in Figure \ref{fig:orthotropic_quantities}, namely \change{(a) $\omega=6.38$\,s$^{-1}$ and (b) $\omega=10.08$\,s$^{-1}$.}}
    \label{fig:orthotropic_field}
\end{figure}

Next, we consider a square orthotropic plate. The kinetic energy spectrum in Figure \ref{fig:orthotropic_spectrum} again has two peaks, which we investigate in Figures \ref{fig:orthotropic_quantities} and \ref{fig:orthotropic_field}. The first resonance is predominantly associated with excitation of the first plate mode, whereas the second resonance is predominantly associated with excitation of the third plate mode. In contrast with the square isotropic plate considered earlier, the second and third modes of the orthotropic plate are not degenerate because symmetry has been broken. In particular, the second mode of the orthotropic plate is antisymmetric with respect to the line $y=b/2$ and hence cannot be excited by the symmetric incident plane wave. This also remains the case with the fourth mode. Thus we observe $c_2=c_4=0$ in Figure \ref{fig:orthotropic_quantities}. We remark that the far field patterns (Figure \ref{fig:orthotropic_quantities}b and d) are symmetric with respect to $\theta=0$, and the surface elevations (Figure \ref{fig:orthotropic_field}) are symmetric about $y=b/2$. 

\begin{figure}
    \centering
    \includegraphics[width=\linewidth]{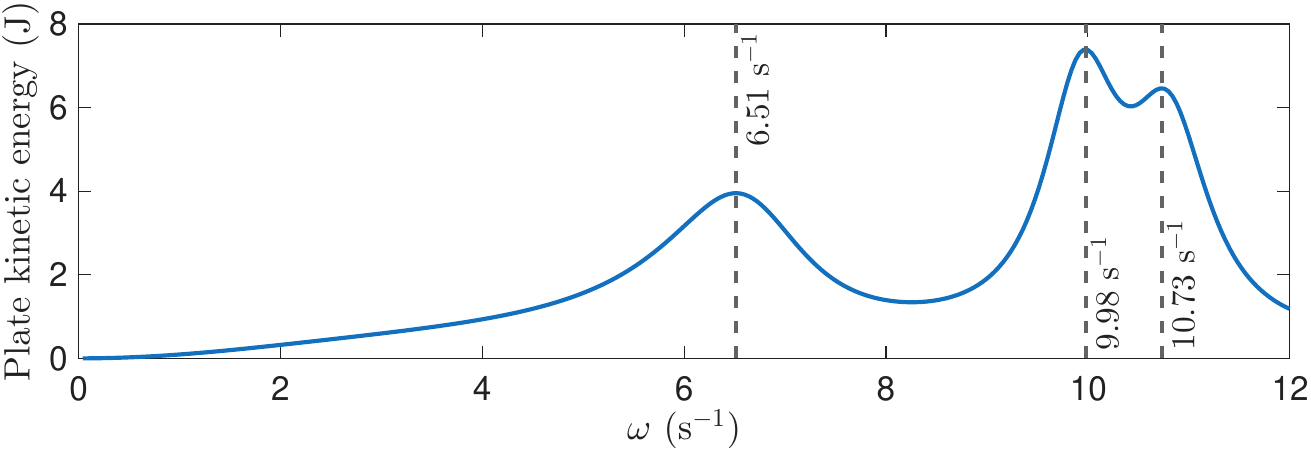}
    \caption{Kinetic energy spectrum for a square anisotropic plate \change{with clamped edges}, with $a=b=1$\,m.}
    \label{fig:anisotropic_spectrum}
\end{figure}

\begin{figure}
    \centering
    \includegraphics[width=\linewidth]{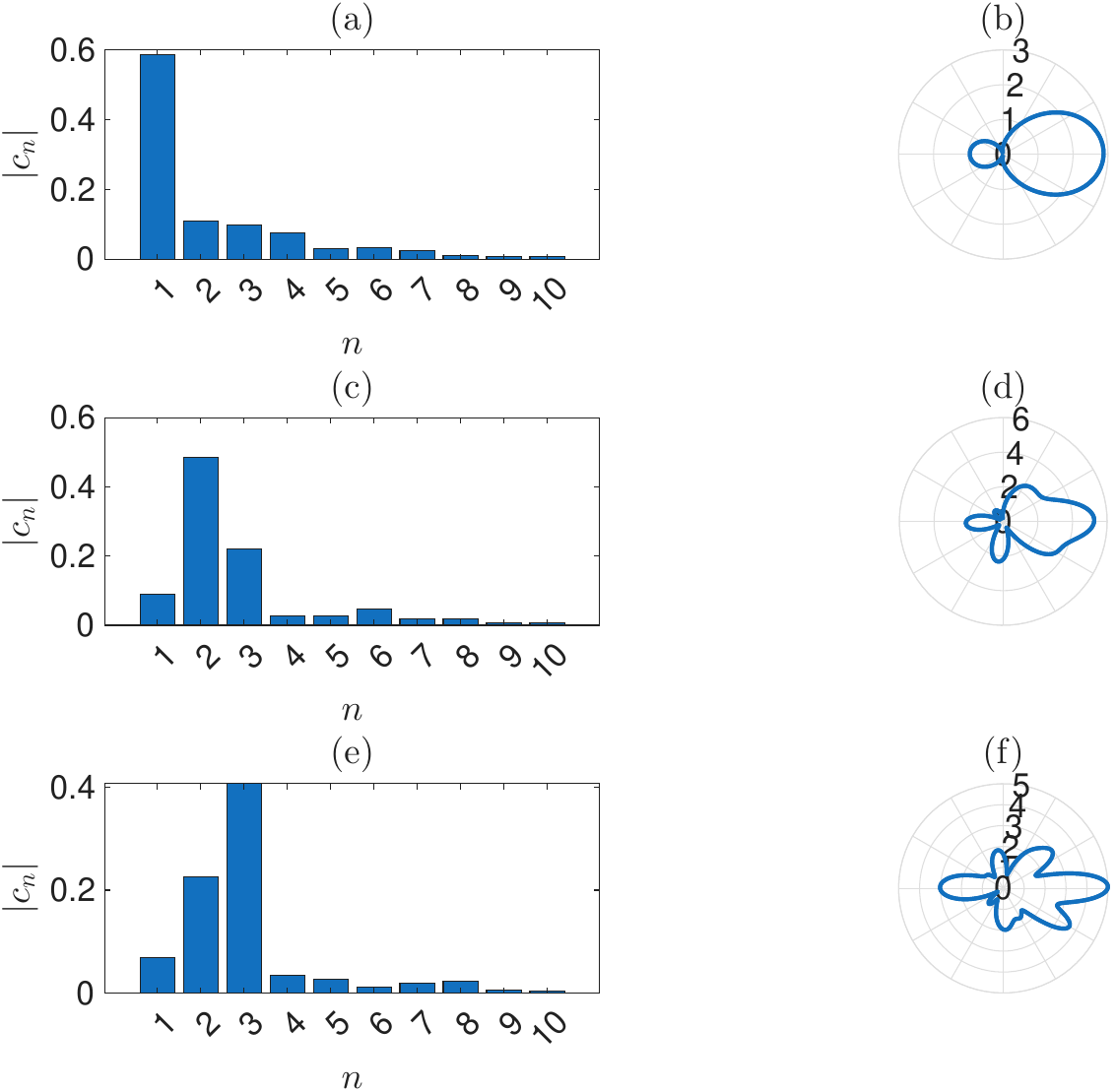}
    \caption{\change{Dry m}ode expansion coefficient magnitudes (left column) and far field patterns (right column) at resonant frequencies of the square anisotropic plate \change{with clamped edges and} with $a=b=1$\,m. The \change{resonant frequencies} are (a,b) \change{$\omega=6.51$\,s$^{-1}$, (c,d) $\omega=9.98$\,s$^{-1}$ and (e,f) $\omega=10.73$\,s$^{-1}$.}}
    \label{fig:anisotropic_quantities}
\end{figure}

\begin{figure}
    \centering
    \includegraphics[width=\linewidth]{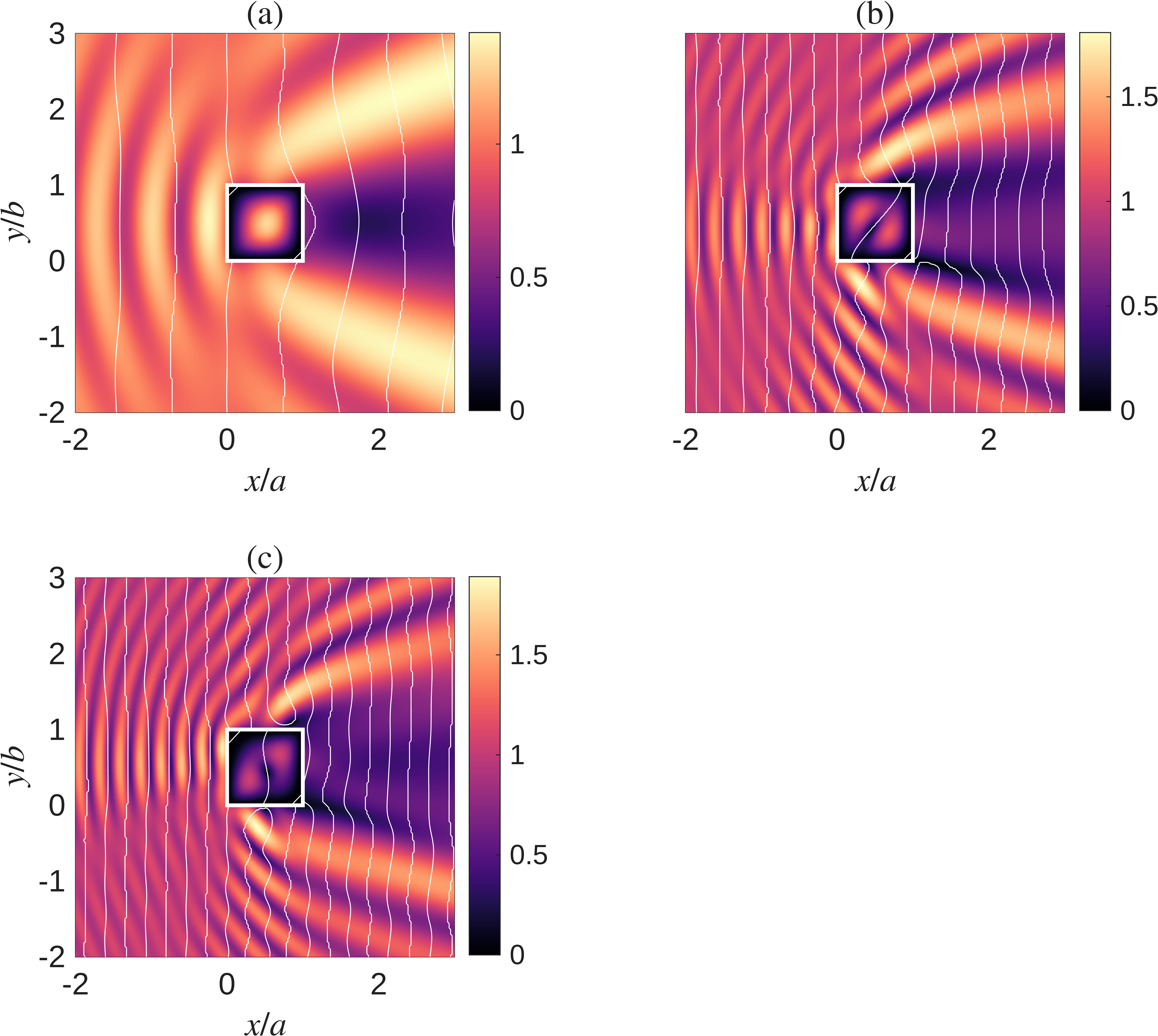}
    \caption{Surface elevation of the excited square \change{an}isotropic plate \change{with clamped edges} ($a=b=1$\,m) at the resonant frequencies considered in Figure \ref{fig:anisotropic_quantities}, namely \change{(a) $\omega=6.51$\,s$^{-1}$, (b) $\omega=9.98$\,s$^{-1}$ and (c) $\omega=10.73$\,s$^{-1}$.}}
    \label{fig:anisotropic_field}
\end{figure}

The next case we consider is a square anisotropic plate, as parametrised by \citet{10.1115/1.4053090}. These parameters, which are reproduced in Table \ref{tab:rigidity_coeffs}, describe a graphite/epoxy laminated plate in which the tensor describing the material is rotated by 45$^\circ$ in the $xy$-plane. As such, the modes of vibration in Figure \ref{fig:mode_shapes} are symmetric or antisymmetric with respect to the lines $y=x$ and $x+y=a$. The response of the surface-mounted square anisotropic plate to a plane wave with incident angle $\chi=0$ is considered in Figures \ref{fig:anisotropic_spectrum}--\ref{fig:anisotropic_field}. We observe three resonant peaks in the kinetic energy spectrum (Figure \ref{fig:anisotropic_spectrum}), which are predominantly associated with excitation of the first three modes of vibration in order of increasing frequency. The second and third modes of vibration are not degenerate, and neither are antisymmetric with respect to the line $y=b/2$, so both can be excited.  We remark that the far field patterns (Figure \ref{fig:anisotropic_quantities}b, d and f) and the surface elevations (Figure \ref{fig:anisotropic_field}) do not possess any symmetry. Observe that while Figures \ref{fig:anisotropic_quantities}b and \ref{fig:anisotropic_field}a may appear at first glance to suggest a symmetric response, this is due to the first mode shape (Figure \ref{fig:mode_shapes}i) being close to symmetric---there are in fact subtle asymmetries in Figure \ref{fig:anisotropic_field}.

\begin{figure}
    \centering
    \includegraphics[width=\linewidth]{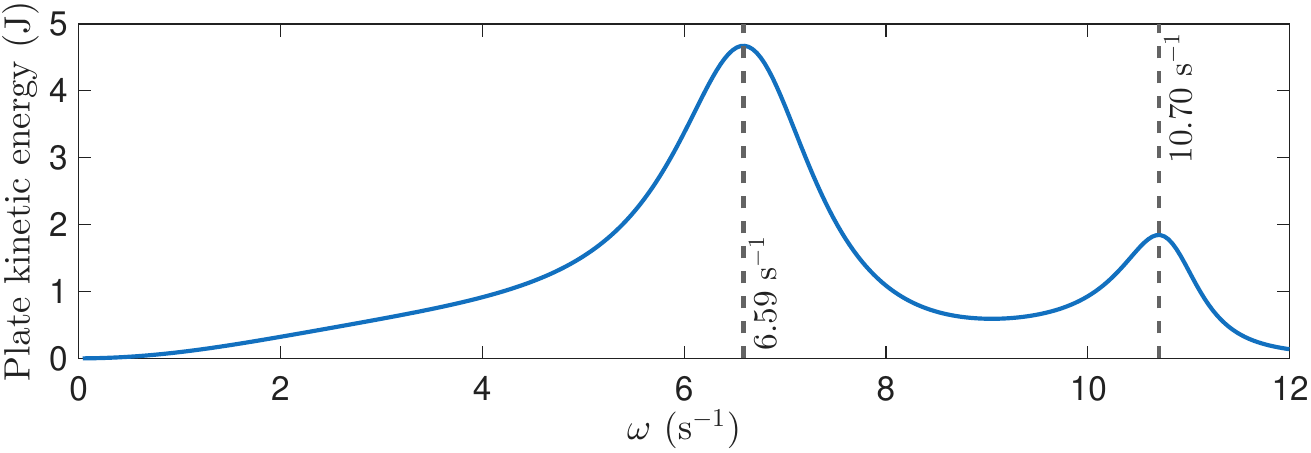}
    \caption{Kinetic energy spectrum for a square anisotropic plate \change{with clamped edges}, with $a=b=1$\,m (as in Figure \ref{fig:anisotropic_spectrum}) and with the direction of the incident wave being $\pi/4$.}
    \label{fig:anisotropic_45_spectrum}
\end{figure}

\begin{figure}
    \centering
    \includegraphics[width=\linewidth]{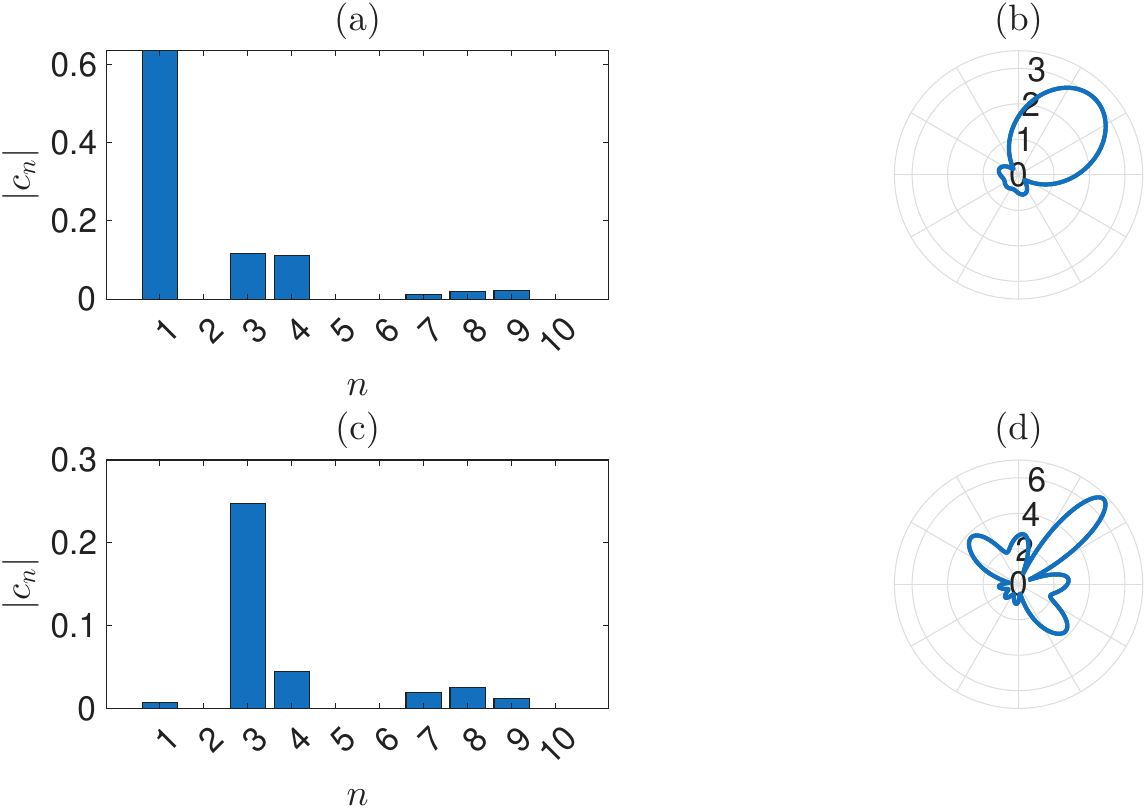}
    \caption{\change{Dry m}ode expansion coefficient magnitudes (left column) and far field patterns (right column) at resonant frequencies of the square anisotropic plate \change{with clamped edges and} with $a=b=1$\,m, with the incident angle $\pi/4$. The \change{resonant frequencies} are (a) \change{$\omega=6.59$\,s$^{-1}$ and (b) $\omega=10.70$\,s$^{-1}$.}}
    \label{fig:anisotropic_45_quantities}
\end{figure}

\begin{figure}
    \centering
    \includegraphics[width=\linewidth]{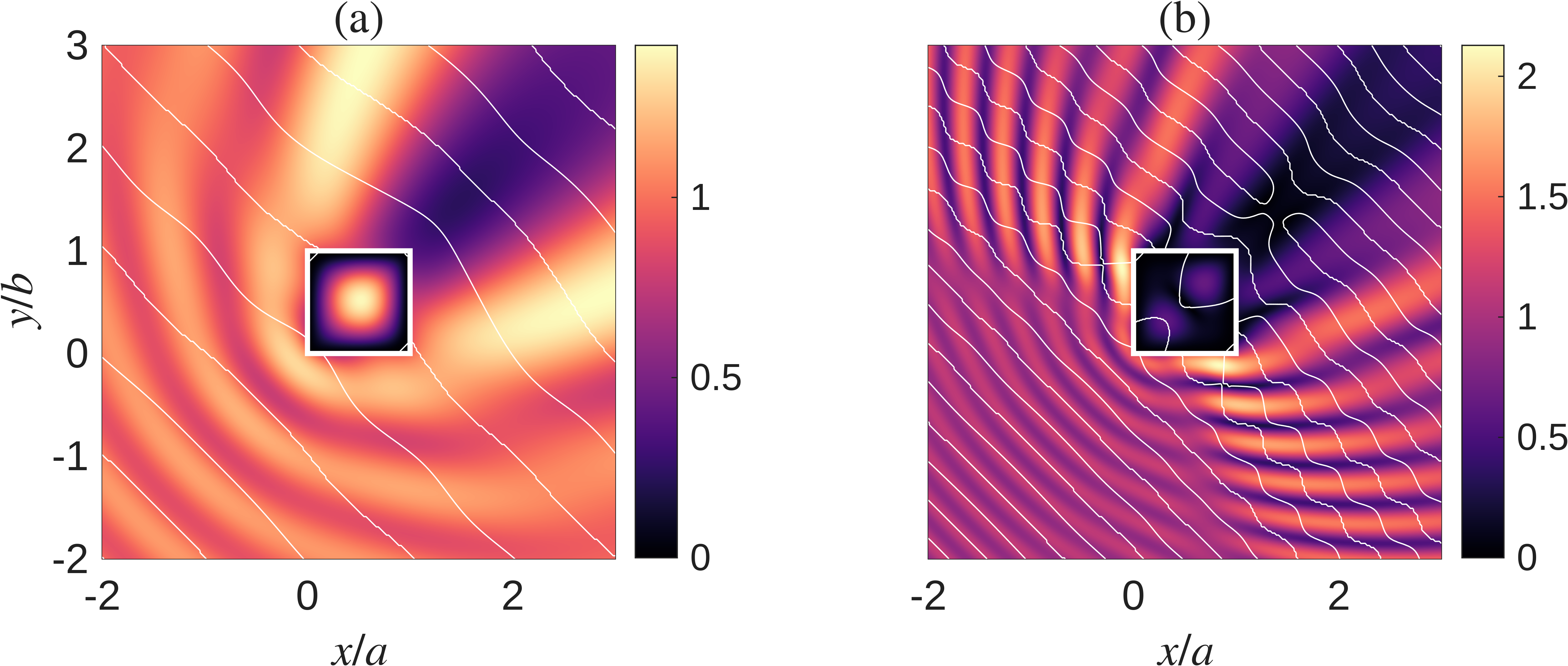}
    \caption{Surface elevation of the excited square \change{an}isotropic plate \change{with clamped edges} ($a=b=1$\,m) with incident angle $\pi/4$ at the resonant frequencies considered in Figure \ref{fig:anisotropic_45_quantities}, namely (a)\change{ $\omega=6.59$\,s$^{-1}$ and (b) $\omega=10.70$\,s$^{-1}$.}}
    \label{fig:anisotropic_45_field}
\end{figure}

Figures \ref{fig:anisotropic_45_spectrum}--\ref{fig:anisotropic_45_field} also explore the square anisotropic plate, although this time the angle of the incident wave is modified to $\chi=\pi/4$. This incident wave is symmetric with respect to the line $y=x$, so modes that are antisymmetric with respect to this line should not resonate. Indeed, we observe that $c_2=0$, \change{i.e.,} the second mode of vibration is not excited because it does have this antisymmetry (see Figure \ref{fig:mode_shapes}). Moreover, there is no peak around \change{$\omega=9.98$\,s$^{-1}$} in the kinetic energy spectrum---\change{which was} associated with excitation of the second mode in the $\chi=0$ case.  We remark that the far field patterns (Figure \ref{fig:anisotropic_45_quantities}b and d) are symmetric with respect to $\theta=\pi/4$, and the surface elevations (Figure \ref{fig:anisotropic_45_field}) are symmetric about the line $y=x$. 

\begin{figure}
    \centering
    \includegraphics[width=\linewidth]{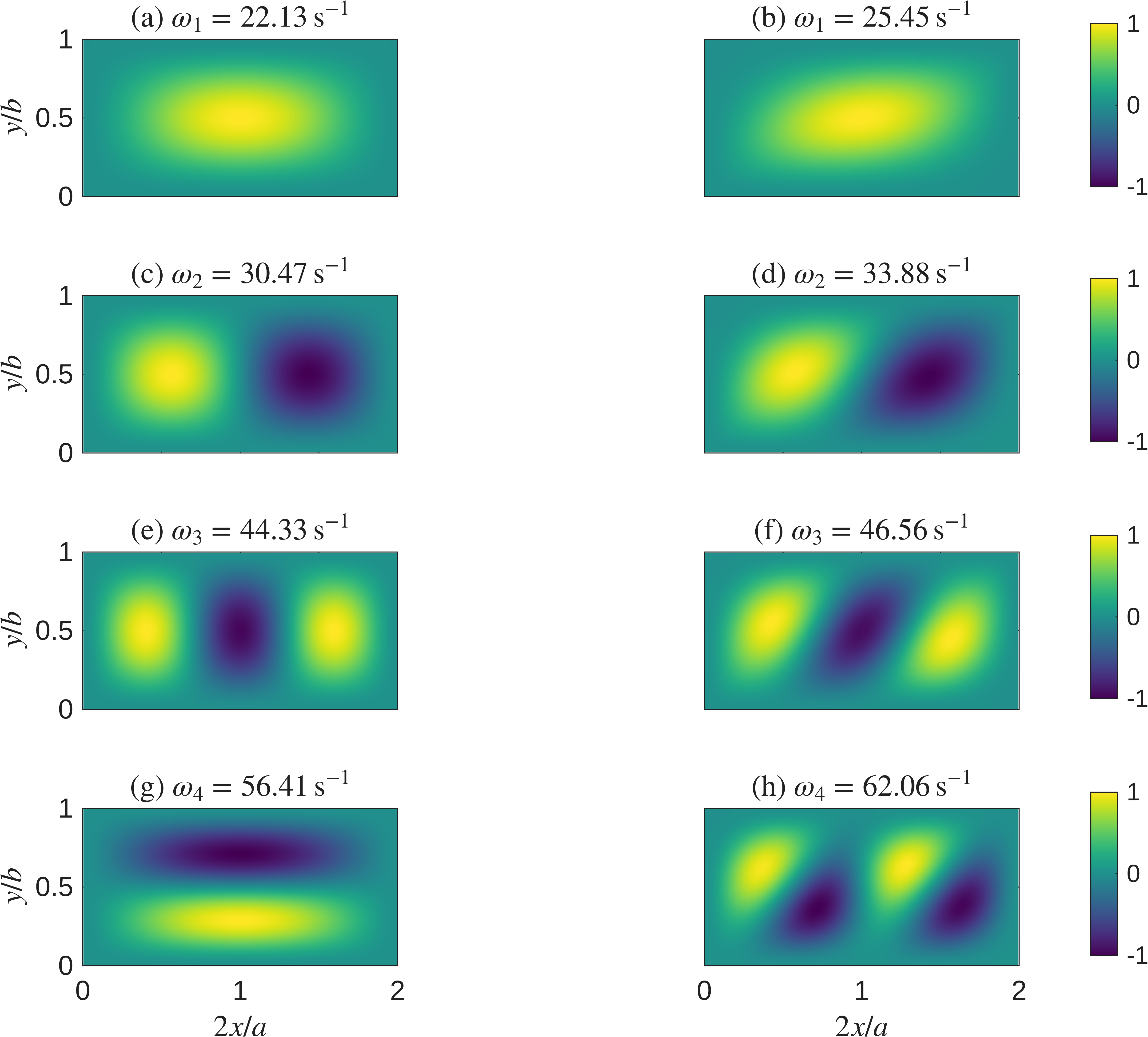}
    \caption{First four mode shapes of an (a,c,e,g) orthotropic and (b,d,f,h) anisotropic rectangular plate, \change{each with clamped edges and} with $a=2b=2$\,m.}
    \label{fig:mode_shapes_rect}
\end{figure}

\begin{figure}
    \centering
    \includegraphics[width=\linewidth]{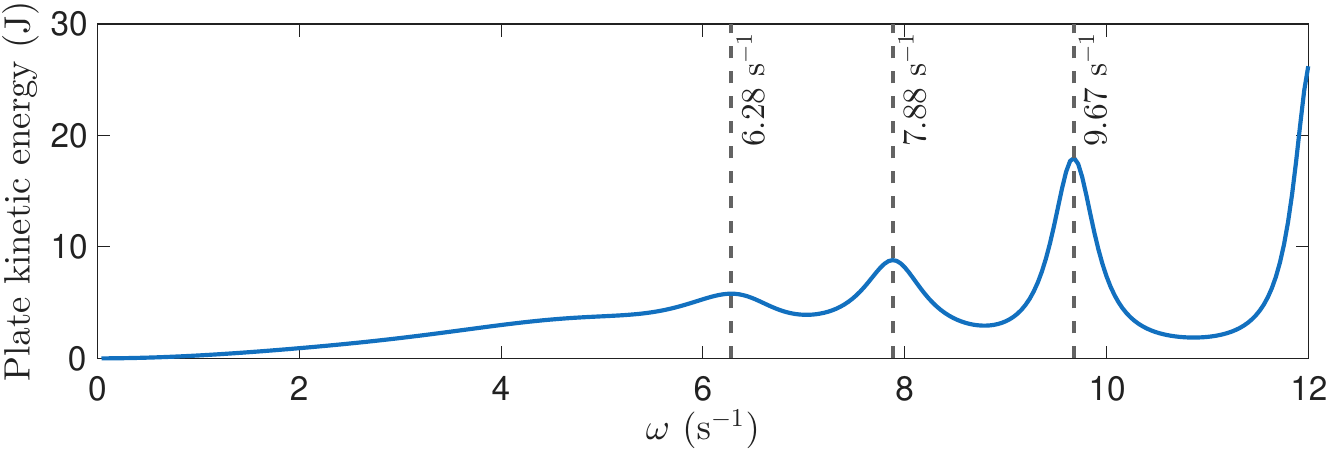}
    \caption{Kinetic energy spectrum for a rectangular orthotropic plate \change{with clamped edges}, with $a=2$\,m and $b=1$\,m.}
    \label{fig:orthotropic_rectangle_spectrum}
\end{figure}

\begin{figure}
    \centering
    \includegraphics[width=\linewidth]{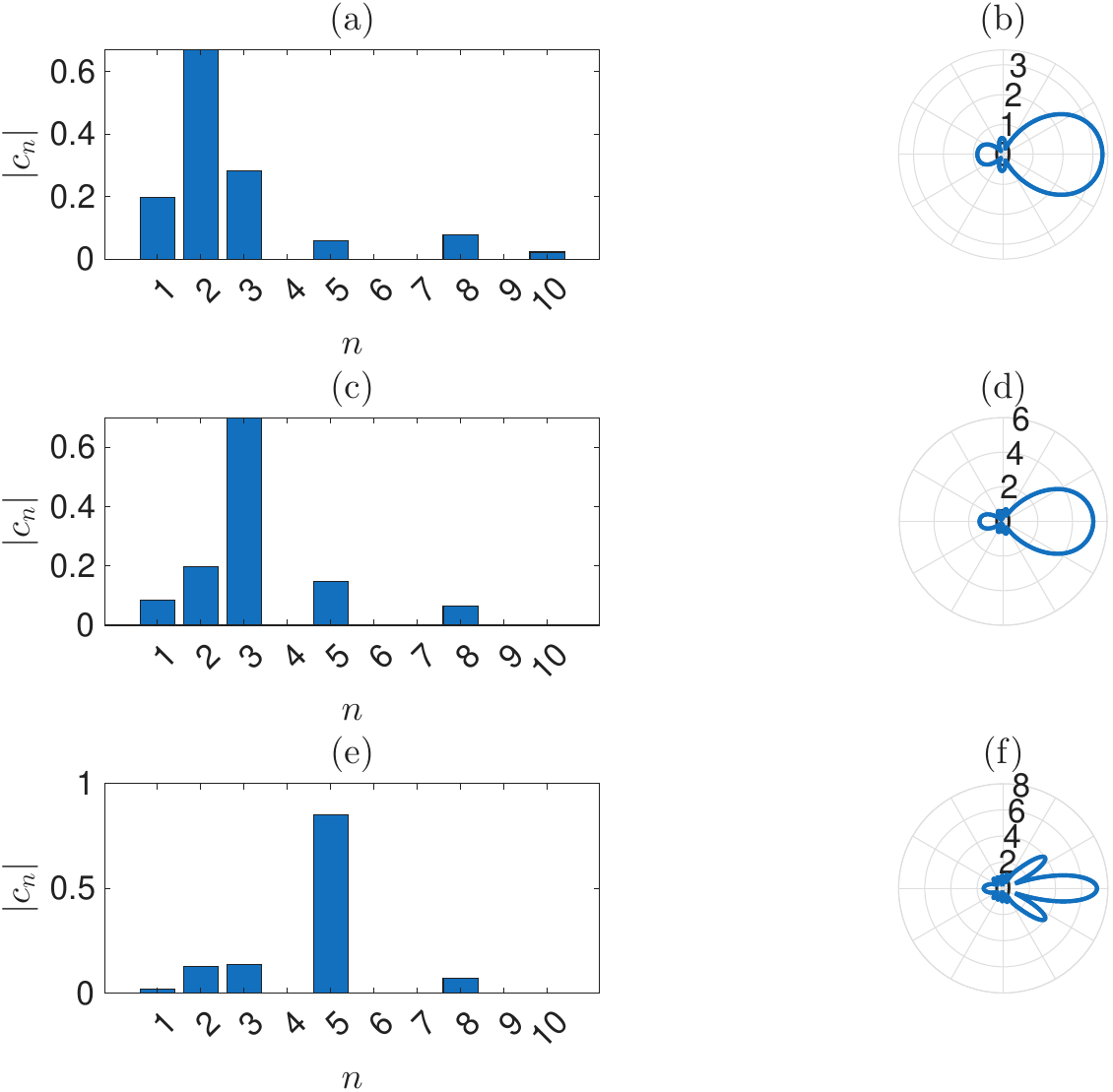}
    \caption{\change{Dry m}ode expansion coefficient magnitudes (left column) and far field patterns (right column) at resonant frequencies of the rectangular orthotropic plate \change{with clamped edges and} with $a=2b=2$\,m. The \change{resonant frequencies are (a,b) $\omega=6.28$\,s$^{-1}$, (c,d) $\omega=7.88$\,s$^{-1}$ and (e,f) $\omega=9.67$\,s$^{-1}$.}}
    \label{fig:orthotropic_rectangle_quantities}
\end{figure}

\begin{figure}
    \centering
    \includegraphics[width=0.8\linewidth]{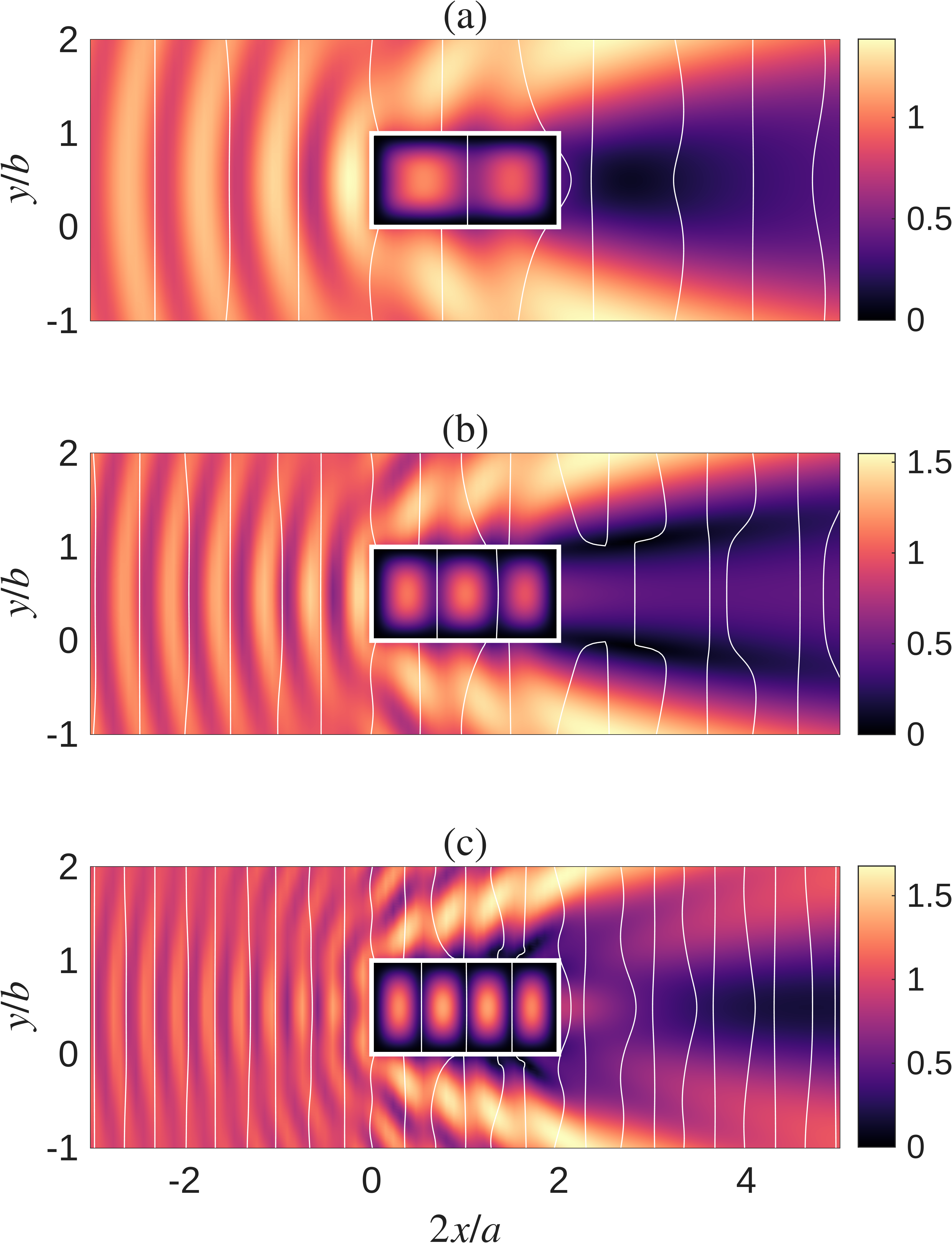}
    \caption{Surface elevation of the excited rectangular orthotropic plate \change{with clamped edges} ($a=2b=2$\,m) at the resonant frequencies considered in Figure \ref{fig:orthotropic_rectangle_quantities}, namely (a)  $\omega=6.28$\,s$^{-1}$, (b) $\omega=7.88$\,s$^{-1}$ and (c) $\omega=9.67$\,s$^{-1}$.}
    \label{fig:orthotropic_rect_field}
\end{figure}

\begin{figure}
    \centering
    \includegraphics[width=\linewidth]{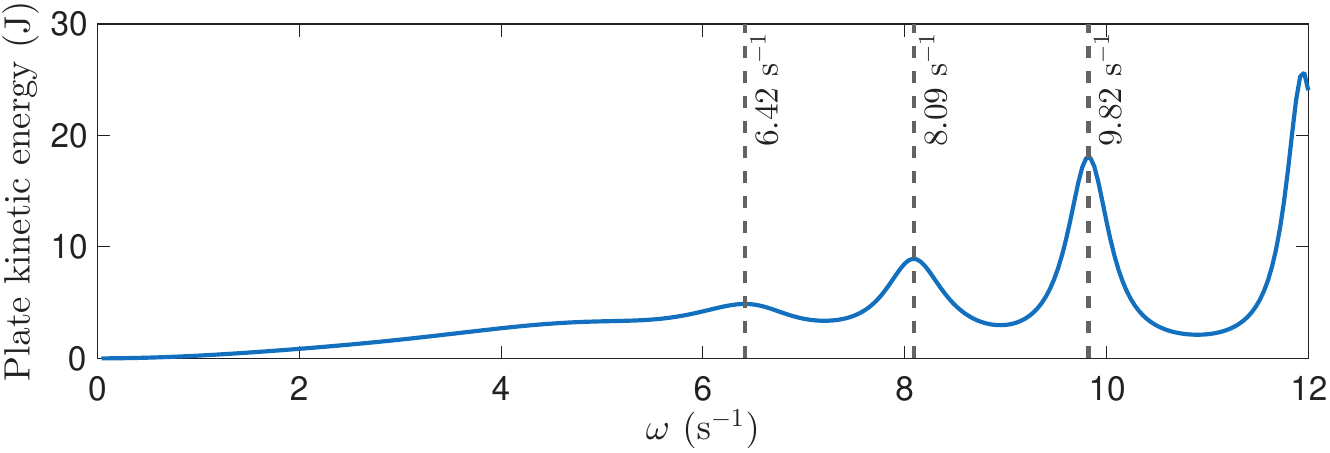}
    \caption{Kinetic energy spectrum for a rectangular anisotropic plate \change{with clamped edges}, with $a=2b=2$\,m.}
    \label{fig:anisotropic_rectangle_spectrum}
\end{figure}

\begin{figure}
    \centering
    \includegraphics[width=\linewidth]{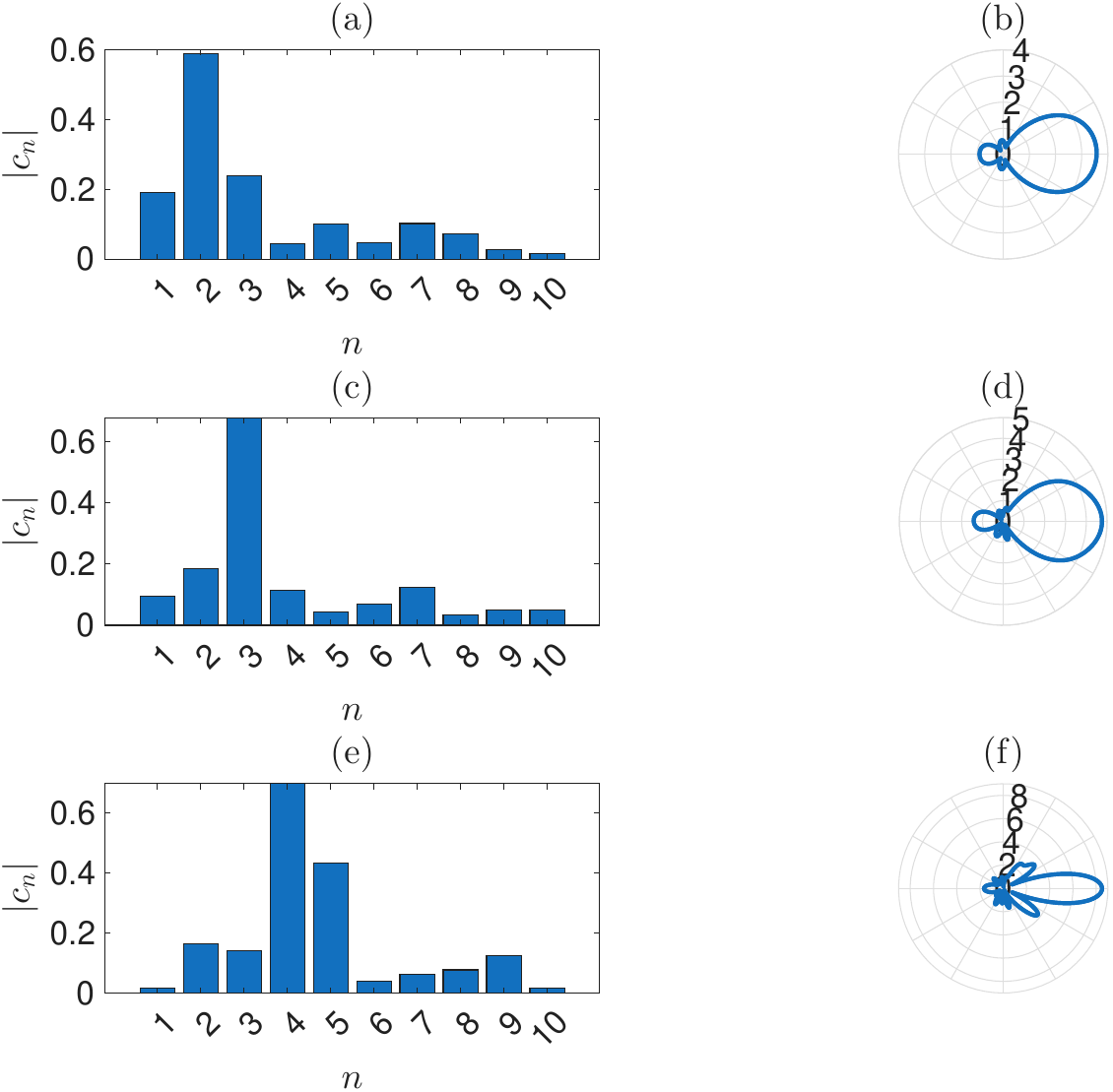}
    \caption{\change{Dry m}ode expansion coefficient magnitudes (left column) and far field patterns (right column) at resonant frequencies of the rectangular anisotropic plate \change{with clamped edges and} with $a=2\,m$ and $b=1$\,m. The \change{resonant frequencies} are (a,b) \change{$\omega=6.42$\,s$^{-1}$, (c,d) $\omega=8.09$\,s$^{-1}$ and (e,f) $\omega=9.82$\,s$^{-1}$.}}
    \label{fig:anisotropic_rectangle_quantities}
\end{figure}

\begin{figure}
    \centering
    \includegraphics[width=0.8\linewidth]{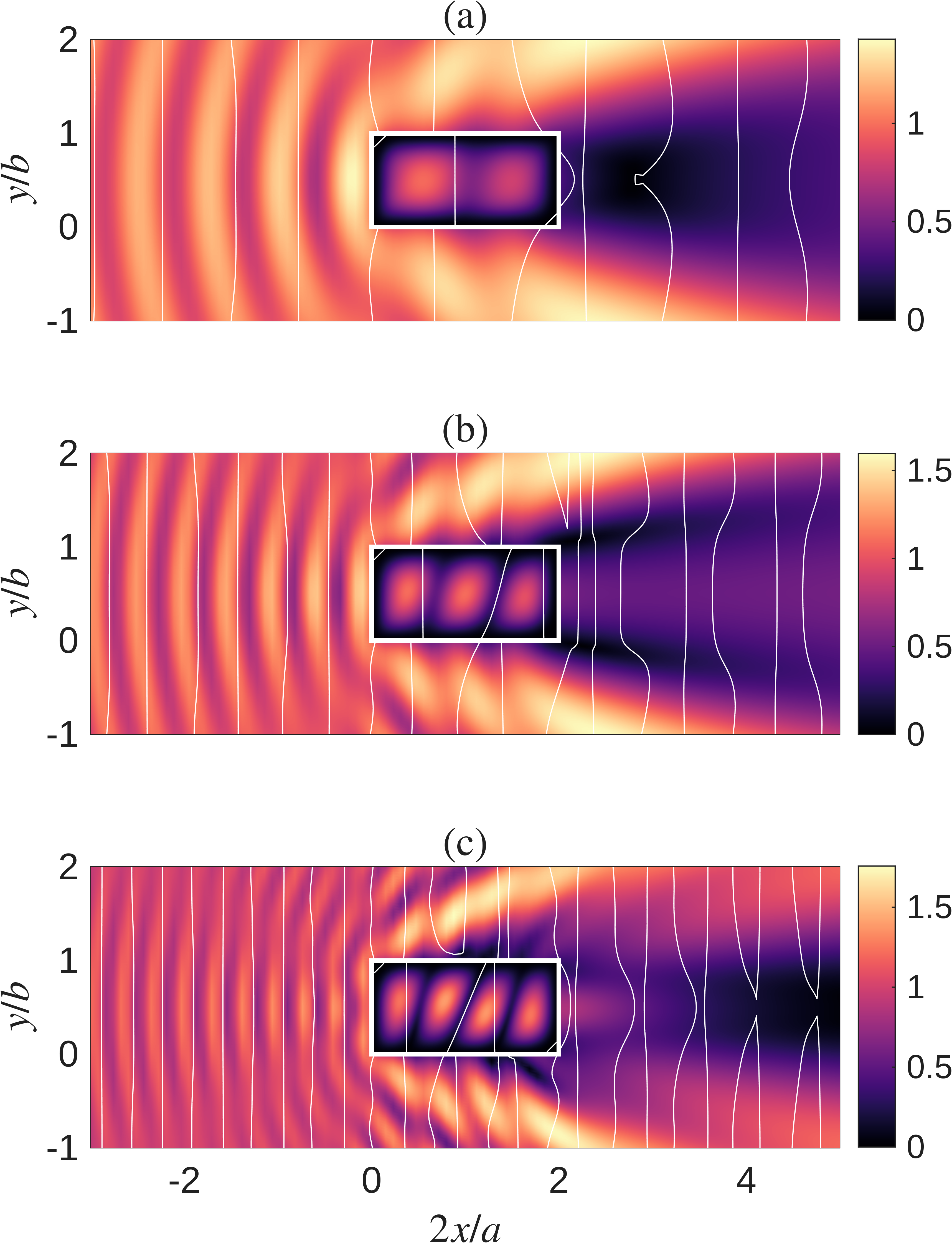}
    \caption{Surface elevation of the excited rectangular anisotropic plate \change{with clamped edges} ($a=2$\,m and $b=1$\,m) at the resonant frequencies considered in Figure \ref{fig:anisotropic_rectangle_quantities}, namely (a)  \change{$\omega=6.42$\,s$^{-1}$, (b)  $\omega=8.09$\,s$^{-1}$ and (c)  $\omega=9.82$\,s$^{-1}$.}}
    \label{fig:anisotropic_rect_field}
\end{figure}

\change{Next}, we present results for rectangular \change{clamped} plates, with $a=2$\,m and $b=1$\,m, considering both the orthotropic and anisotropic cases. The first four mode shapes of both cases are given in Figure \ref{fig:mode_shapes_rect}. Their kinetic energy spectra are given in Figures \ref{fig:orthotropic_rectangle_spectrum} and \ref{fig:anisotropic_rectangle_spectrum}, each having three resonant peaks in the frequency interval considered. The relative excitations of the various plate modes at these resonant frequencies are given in Figure \ref{fig:orthotropic_rectangle_quantities} for the orthotropic case, and Figure \ref{fig:anisotropic_rectangle_quantities} for the anisotropic case. In both cases, we do not observe a strong resonant peak associated with the first plate mode. Moreover, in the orthotropic case, the fourth resonant mode is never excited because it is antisymmetric with respect to $y=b/2$, whereas the incident wave is symmetric with respect to this line. Finally, surface elevations of the first three resonances of orthotropic and anisotropic plates are given in Figures \ref{fig:orthotropic_rect_field} and \ref{fig:anisotropic_rect_field}, respectively. 

\begin{figure}
    \centering
    \includegraphics[width=\textwidth]{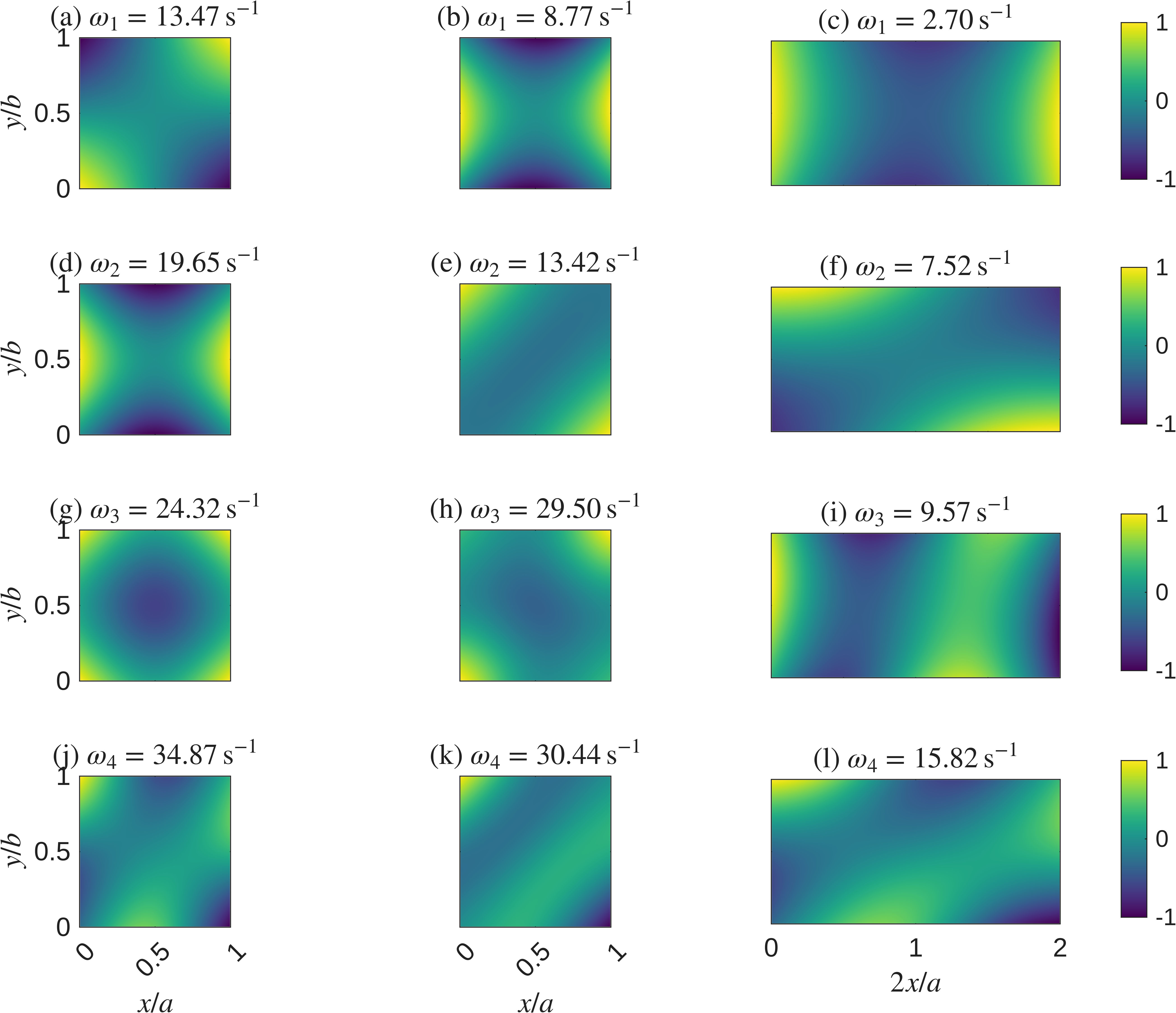}
    \caption{\change{First four non-rigid mode shapes and frequencies of vibration of plates with free edges. Panels (a,d,g,j) show the modes of a square isotropic plate ($a=b=1$\,m), (b,e,h,k) show the modes of a square anisotropic plate ($a=b=1$\,m) and (c,f,i,l) show the modes of a rectangular anisotropic plate ($a=2b=2$\,m). We note that in each case, the first three modes, which are rigid body motions with frequency of vibration $0$\,s$^{-1}$, are not shown.}}
    \label{fig:free_plate_modes}
\end{figure}

\begin{figure}
    \centering
    \includegraphics[width=\linewidth]{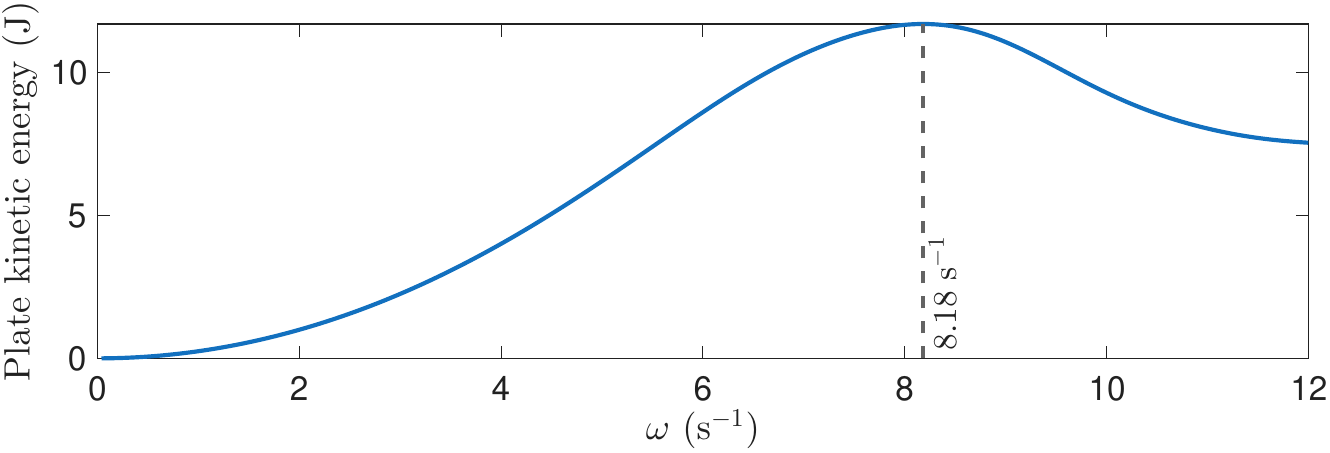}
    \caption{\change{Kinetic energy spectrum for a square anisotropic plate with free edges, with $a=b=1$\,m.}}
    \label{fig:square_spectrum_free}
\end{figure}

\begin{figure}
    \centering
    \includegraphics[width=\linewidth]{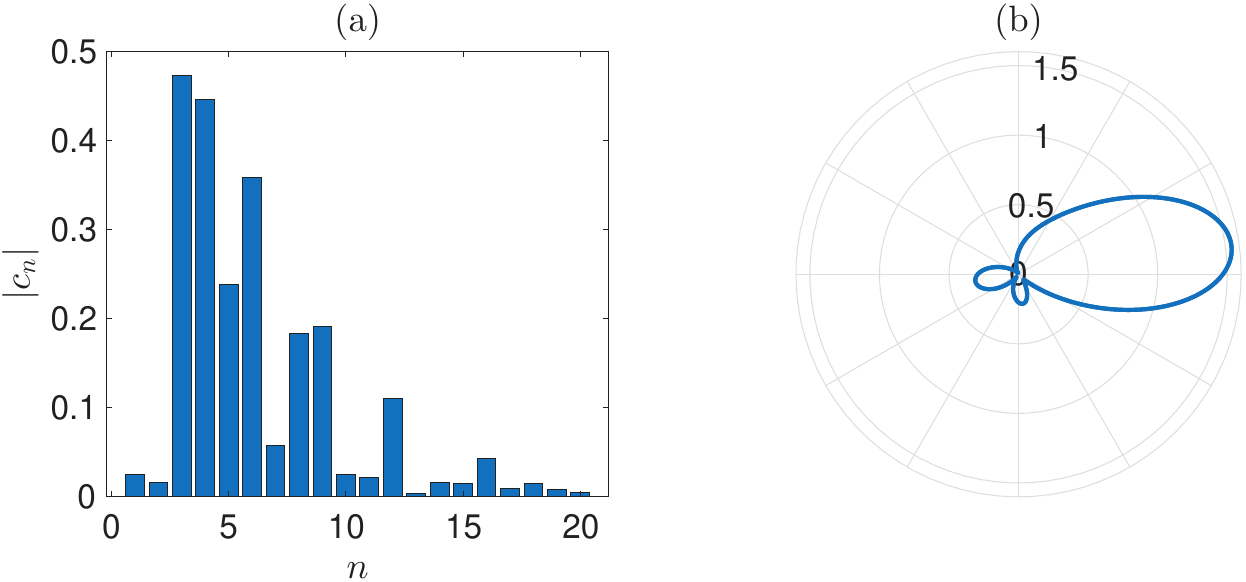}
    \caption{\change{Dry mode expansion coefficient magnitudes (a) and far field patterns (b) at the resonant frequency $\omega=8.18$\,s$^{-1}$ of the square anisotropic plate with free edges, with $a=b=1\,$m. }}
    \label{fig:square_quantities_free}
\end{figure}

\begin{figure}
    \centering
    \includegraphics[width=0.6\linewidth]{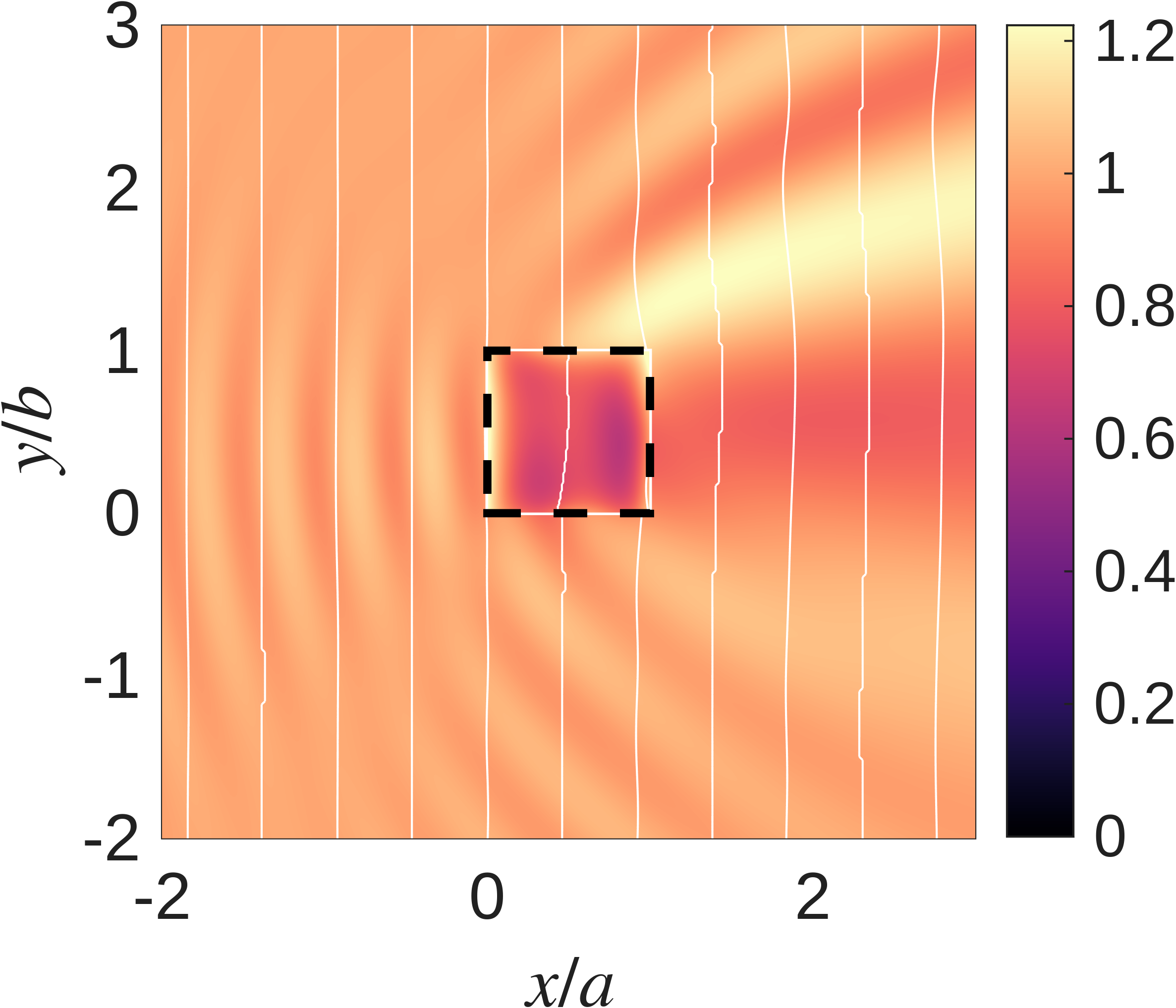}
    \caption{\change{Surface elevation of the excited square anisotropic plate with free edges ($a=b=1$\,m) at the resonant frequency $\omega=8.18$\,s$^{-1}$. In contrast with earlier figures, the free boundary is marked with a black dashed line.}}
    \label{fig:square_field_free}
\end{figure}

\begin{figure}
    \centering
    \includegraphics[width=\linewidth]{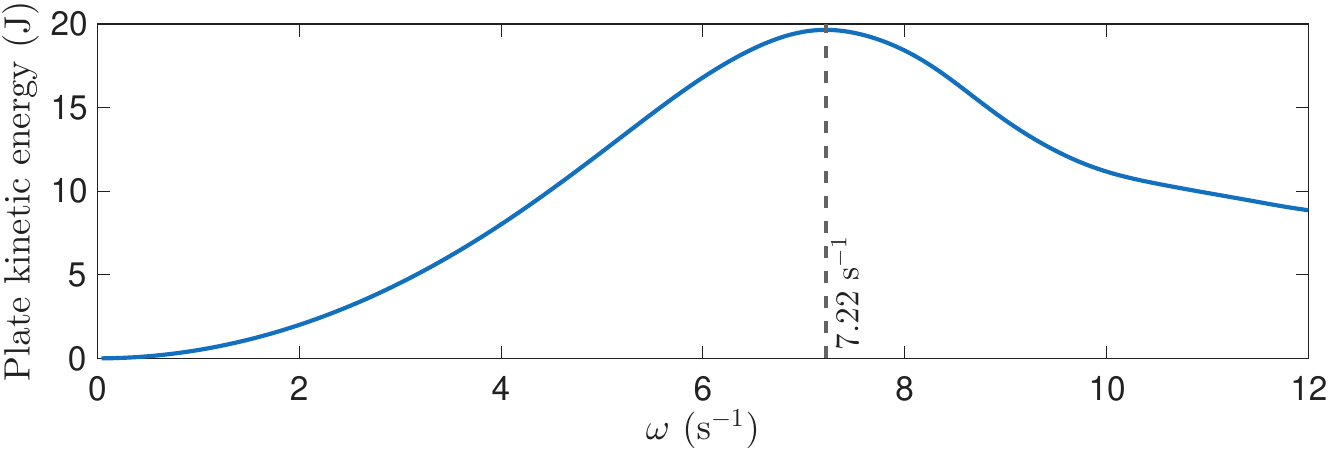}
    \caption{\change{Kinetic energy spectrum for a rectangular anisotropic plate with free edges, with $a=2b=2$\,m.}}
    \label{fig:rect_spectrum_free}
\end{figure}

\begin{figure}
    \centering
    \includegraphics[width=\linewidth]{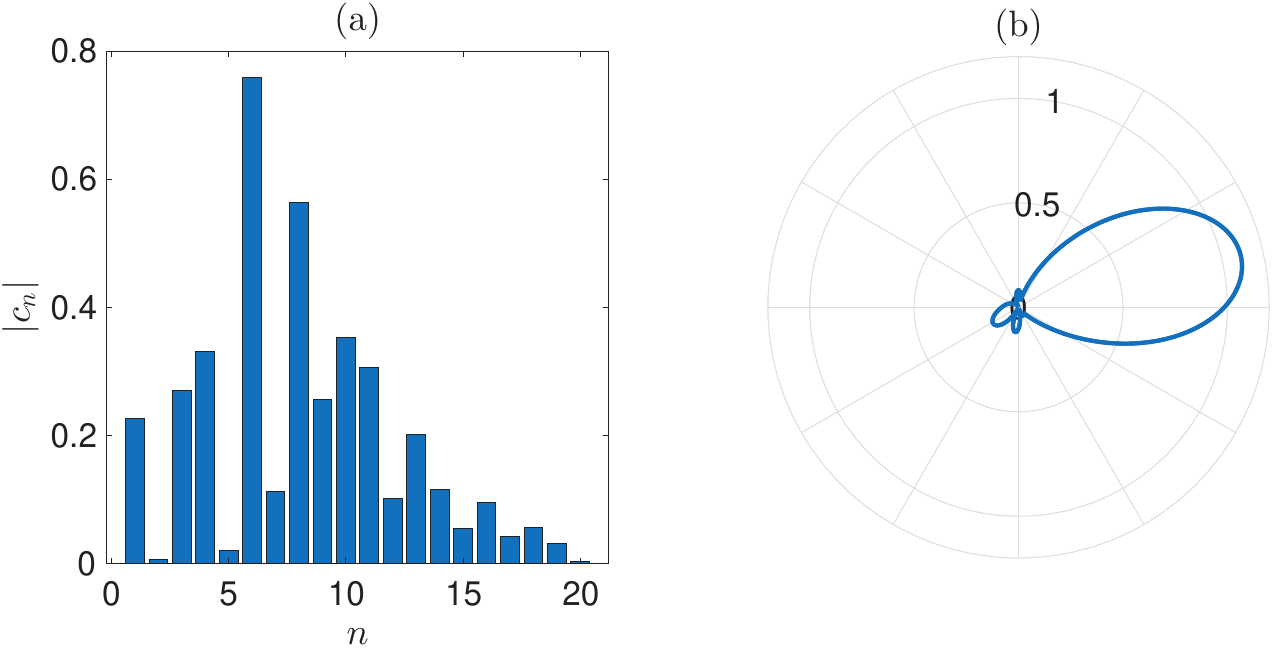}
    \caption{\change{Dry mode expansion coefficient magnitudes (a) and far field patterns (b) at the resonant frequency $\omega=7.22$\,s$^{-1}$ of the rectangular anisotropic plate with free edges, with $a=2b=2\,$m. }}
    \label{fig:rect_quantities_free}
\end{figure}

\begin{figure}
    \centering
    \includegraphics[width=\linewidth]{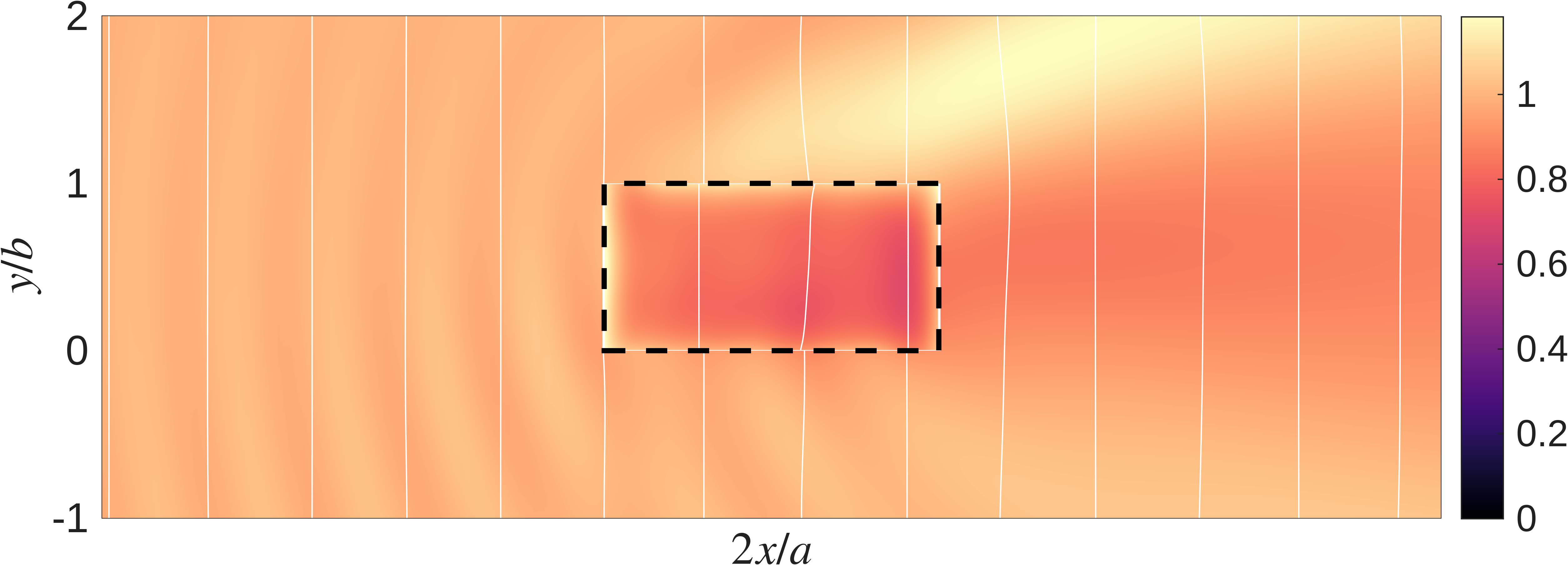}
    \caption{\change{Surface elevation of the excited rectangular anisotropic plate with free edges ($a=2b=2$\,m) at the resonant frequency $\omega=7.22$\,s$^{-1}$.}}
    \label{fig:rect_field_free}
\end{figure}

\change{Next, we consider anisotropic plates with free edges, in both square ($a=b=1$\,m) and rectangular ($a=2b=2$\,m) geometries. Modes of vibration are illustrated in Figure \ref{fig:free_plate_modes}, which also includes the case of a square isotropic plate for the purpose of validation against \citet[][Table C15]{LEISSA1973257}. Kinetic energy spectrums for square and rectangular anisotropic plates with free edges are given in Figures \ref{fig:square_spectrum_free} and \ref{fig:rect_spectrum_free}, respectively. Each of these has only one resonant peak in the frequency interval considered, which is noticeably broader than peaks in the corresponding spectra for clamped plates (Figures \ref{fig:anisotropic_spectrum} and \ref{fig:anisotropic_rectangle_spectrum}, respectively). Figures \ref{fig:square_quantities_free} and \ref{fig:rect_quantities_free} show the dry mode coefficient amplitudes and far field patterns at the resonant frequency. We observe that significant excitation occurs over many modes. In contrast, corresponding figures for clamped plates (Figures \ref{fig:anisotropic_quantities} and \ref{fig:anisotropic_rectangle_quantities}) show that the resonances are predominantly due to excitation of a single mode. Figures \ref{fig:square_field_free} and \ref{fig:rect_field_free} show the surface elevations for square and rectangular anisotropic plates with free edges, respectively.}

\begin{figure}
    \centering
    \includegraphics[width=\textwidth]{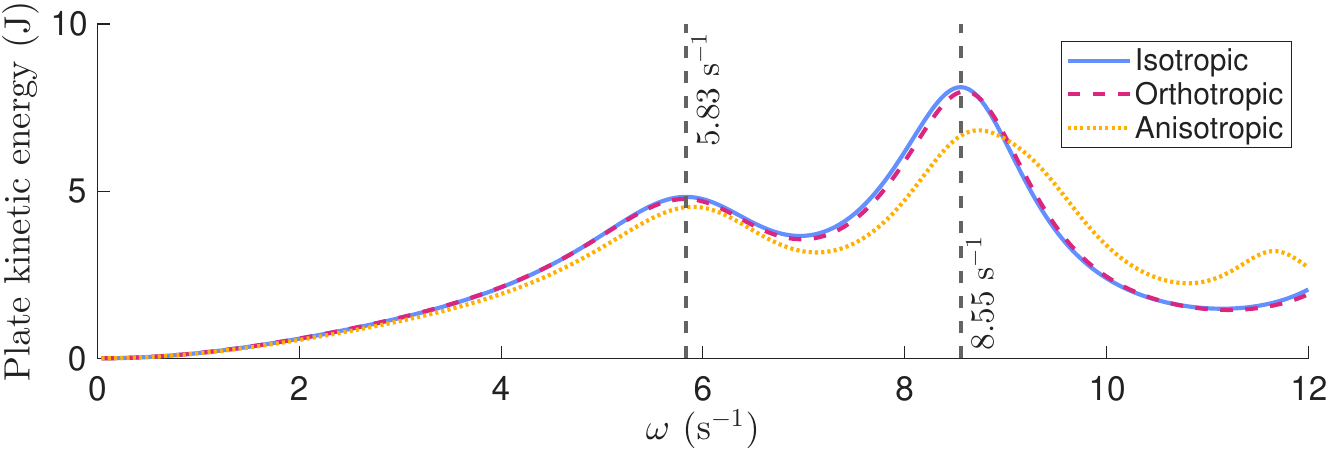}
    \caption{\ctwo{Kinetic energy spectra for isotropic, orthotropic and anisotropic square plates ($a=b=1$\,m) with simply-supported edges. Only the resonances of the isotropic case are marked.}}
    \label{fig:SS_spectra}
\end{figure}

\ctwo{Finally, in Figure \ref{fig:SS_spectra} we present kinetic energy spectra for plates with simply supported edges. While this is primarily included to serve as a benchmark calculation, we note that the resonant frequencies are lower than those of clamped plates (e.g., compare Figures \ref{fig:isotropic_spectrum} and \ref{fig:SS_spectra} for isotropic plates), owing to their lower frequencies of vibration \citep{LEISSA1973257}.} 

\section{Conclusion}\label{conclusion_sec}
This paper has considered the problem of water wave scattering by a surface mounted rectangular anisotropic elastic plate, with \change{either clamped or free} boundaries. The problem was solved by expanding in the modes of vibration of the plate, with the required diffraction and radiation problems being solved using a boundary integral equation/constant panel method. Results for isotropic, orthotropic and anisotropic plates of either square or rectangular geometries were presented that illustrate the capabilities of the numerical method. Overall, our results show that the response by the plate can depend greatly on the rigidity coefficients $D_{ij}$. Our investigation has emphasised that the excitation of certain modes can be forbidden if they are symmetrically opposed to the incident wave.

The main motivation of this work was an application to the modelling of PWECs, and this remains as an area of future work. The method presented here can be used (essentially verbatim) for a surface mounted rectangular piezoelectric plate, with the rigidity coefficients $D_{ij}$ becoming complex. The only necessary adjustment resulting from this would be to the \change{dry mode} expansion method presented in \textsection\ref{coupled_sec}, because the set of plate eigenmodes $w_j$ \change{may no longer be orthogonal in general, which would cause the hydrodynamic matrices $K_{\rm stiff}$, $M$ and $C$ to have non-zero off diagonal entries proportional to $\iint_{\Gamma\times\{0\}} w_jw_m^*\,\upd S$. With this detail taken into account, we} stress that our approach will allow the anisotropy of the piezoelectric material within the WEC to be modelled appropriately in three dimensions. \change{We note that PWECs have not been considered in this paper because the determination of the complex rigidity coefficients from the poling direction, circuiting conditions, and layering of the piezoelectric material is a complicated modelling problem \citep{renzi2016hydroelectromechanical}.} Another possible extension of this work is to arbitrary shaped plates, as considered by \cite{meylan2002wave} for isotropic plates with free edges. Although the modes of vibration would need to be calculated using a different method to the one presented in \textsection\ref{vibration_modes_sec} (e.g., the finite element method could be used), the hydrodynamic component of the solution could remain unchanged. Lastly, we mention the problem of extending this study to submerged horizontal plates. This case is more complicated from a mathematical perspective because it leads to hypersingular integral equations.

\subsection*{Funding}
This study was supported by the Australian Research Council's Discovery Projects funding scheme (DP240102104).

\subsection*{Declaration of Interests}
The authors report no conflict of interest.

\subsection*{Data availability}
Data sharing does not apply to this article, as no new data was created or analysed in this study. The computer code used to generate the solutions is available from the corresponding author upon reasonable request.

\appendix


\change{\section{Derivation of the Green's function}\label{Greens_derivation}
In this appendix, we give a derivation of the Green's function introduced in \textsection\ref{greens_fn_subsec}. The equation satisfied by the Green's function \eqref{Greens_fn_pde}, when written in radially symmetric cylindrical coordinates, becomes
\begin{subequations}
\begin{align}
\partial_r^2 G(r,z)+\frac{1}{r}\partial_rG(r,z)+\partial_z^2 G(r,z)&=0&r> 0,\,-H<z<0,\label{laplace_greens}\\
\partial_zG(r,z)&=0&z=-H,\label{sb_greens}\\
\partial_zG(r,z)-\alpha G(r,z)&=-\frac{1}{2\pi r}\delta(r)&z=0,\label{fs_greens}\\
\sqrt{r}(\partial_r-\upi k)G(r,z)&\to0&r\to\infty.\label{sommerfeld_greens}
\end{align}
\end{subequations}
Let $\hat{G}$ be the zeroth Hankel transform of $G$, i.e.
\begin{equation}
\hat{G}(\mu,z)=\int_0^\infty G(r,z) \mathrm{J}_0(\mu r)r\upd r,
\end{equation}
where $\mathrm{J}_0$ is the Bessel function of the first kind of order zero. The inverse transform is given by
\begin{equation}\label{Inverse_hankel}
G(r,z)=\int_0^\infty \hat{G}(\mu,z) \mathrm{J}_0(\mu r)\mu\upd\mu.
\end{equation}
After transforming \eqref{laplace_greens} and \eqref{sb_greens}, we obtain
\begin{subequations}
\begin{align}
-\mu^2\hat{G}(\mu,z)+\partial_z^2\hat{G}(\mu,z)&=0&\mu>0,\,-H<z<0,\\
\partial_z\hat{G}(\mu,z)&=0&z=-H,
\end{align}
\end{subequations}
whereby
\begin{equation}\label{G_hat}
\hat{G}(\mu,z)=A(\mu)\cosh\mu(z+H).
\end{equation}
The function $A(\mu)$ must be determined from the boundary condition at $z=0$ \eqref{fs_greens}. The Hankel transform of this equation is
\begin{equation}
\partial_z\hat{G}-\alpha\hat{G}=-\frac{1}{2\pi},
\end{equation}
\ctwo{for $z=0$<} so we compute
\begin{align}
A(\mu)&=-\frac{1}{2\pi}\frac{1}{\mu\sinh\mu H-\alpha\cosh \mu H}\nonumber\\
&=-\frac{\e^{-\mu H}}{2\pi\mu}\left(1+\frac{\mu+\alpha}{\mu\sinh\mu H-\alpha\cosh\mu H}\right),\label{A_of_mu}
\end{align}
where the second line follows from a substantial sequence of algebraic manipulations. Taking the inverse Hankel transform of $\hat{G}$ \eqref{Inverse_hankel}, with consideration of \eqref{G_hat} and \eqref{A_of_mu}, yields
\begin{align}
G(r,z)&=-\int_0^\infty \frac{\e^{-\mu H}}{2\pi\mu}\cosh\mu(z+H)\mathrm{J}_0(\mu r)\mu\upd\mu\nonumber\\
&\qquad-\Uint_0^\infty\frac{\e^{-\mu H}}{2\pi\mu}\frac{(\mu+\alpha)\cosh\mu H}{\mu\sinh\mu H-\alpha\cosh\mu H}\cosh\mu(z+H)\mathrm{J}_0(\mu r)\mu\upd\mu.
\end{align}
Note that we must integrate beneath the pole of the second integrand in order to satisfy the Sommerfeld radiation condition \eqref{sommerfeld_greens}. After expressing $\cosh\mu(z+H)=\tfrac{1}{2}(\e^{\mu(z+H)}+\e^{-\mu(z+H)})$ and using the fact that \citep{Bracewell1999}
\begin{equation}
\int_0^{\infty}\frac{1}{\sqrt{r^2+\zeta^2}}\mathrm{J}_0(\mu r) r\upd r=\frac{\e^{-\mu|\zeta|}}{\mu},
\end{equation}
the first integral can be evaluated, yielding \eqref{greens1}.}

\section{Beam eigenfunctions}\label{beam_eigenfunction_sec}
The solutions of the eigenfunction problem associated with a beam clamped at both ends \eqref{beam_equation} may be expressed as
\begin{equation}\label{beam_mode_def}
    u(x)=a_1\sin(\kappa x)+a_2\cos(\kappa x)+a_3\e^{-\kappa x}+a_4\e^{\kappa(x-1)}.
\end{equation}
In order for the boundary conditions to be satisfied, we must find $\kappa$ for which the following nullspace problem has a nontrivial solution
\begin{equation}\label{beam_matrix}
    \begin{bmatrix}
        0&1&1&\e^{-\kappa }\\
        \sin(\kappa )&\cos(\kappa )&\e^{-\kappa}&1\\
        1&0&-1&\e^{-\kappa }\\
        \cos(\kappa )&-\sin(\kappa )&-\e^{-\kappa}&1
    \end{bmatrix}\begin{bmatrix}
        a_1\\a_2\\a_3\\a_4
    \end{bmatrix}=\begin{bmatrix}
        0\\0\\0\\0
    \end{bmatrix}.
\end{equation}
This is true when the determinant of the matrix is zero, which can be shown to occur when
\begin{equation}
    \sech(\kappa)=\cos(\kappa).
\end{equation}
\change{We note that the clamped-clamped beam does not support modes for $\kappa=0$.}

\change{In the case that the beam is free at both ends, as in equation \eqref{beam_equation_free}, the ansatz \eqref{beam_mode_def}, \ctwo{upon formation of a matrix analogous to that in \eqref{beam_matrix},} gives rise to the same transcendental equation for the beam eigenvalues $\kappa_m$ as the clamped case, namely
\begin{equation}
    \sech(\kappa)=\cos(\kappa).
\end{equation}
For non-zero roots of the above, the corresponding eigenmodes are of the form \eqref{beam_mode_def}. Note that the free-free beam also supports two modes for $\kappa_1=\kappa_2=0$, namely
\begin{subequations}
\begin{align}
u_1(x)&=1,\\
u_2(x)&=2\sqrt{3}(x-\tfrac{1}{2}),
\end{align}
\end{subequations}
which correspond to rigid body motions. The remaining modes $u_j$ for $j\geq 3$ are of the form \eqref{beam_mode_def}.}

\ctwo{The eigenvalues and eigenmodes for the simply supported case \eqref{beam_equation_SS} have the following simple form:
\begin{subequations}
\begin{align}
    \kappa_n&=n\pi,\\
    u_n(x)&=\sqrt{2}\sin(n\pi).
\end{align}
\end{subequations}}

We remark that the tasks of normalising of the eigenfunctions $u_m$ in accordance with \eqref{beam_modes_orthonormality}, and computing the quantities \change{$\mathcal{U}_{mp}^{[j,l]}$} introduced in \eqref{U_inner_product},  require the computation of integrals that would be tedious to calculate by hand. Instead, symbolic computation was used to obtain suitable expressions, which were then verified by numerical integration.

\bibliographystyle{jfm}
\bibliography{main}

\end{document}